\documentclass[aps,pre,reprint,twocolumn,superscriptaddress,groupaddress,floatfix]{revtex4-1}

\usepackage[utf8]{inputenc}
\usepackage{graphicx}
\usepackage{bm}
\usepackage{amsmath,mathtools}
\usepackage{amsfonts}
\usepackage{amssymb}
\usepackage{graphicx}
\usepackage[dvipsnames]{xcolor}
\usepackage{color}
\usepackage{booktabs}
\usepackage{multirow}
\usepackage{stackrel}
\newcommand{\vt}[2]{\vartheta_{#1}^{#2}}
\newcommand{\pdt}{\frac{\partial}{\partial t}}
\newcommand{\pdx}{\frac{\partial}{\partial x}}
\begin{document}

\title{Run-and-tumble dynamics with non-reciprocal transitions between three velocity states}

\author{Julio C. R. Romo-Cruz}
\affiliation{Instituto de Física, Universidad Nacional Autónoma de México,
Apartado Postal 20-364, Ciudad de México, México.}

\author{Francisco J. Sevilla}
\email[]{fjsevilla@fisica.unam.mx}
\thanks{author to whom correspondence should be addressed.}
\affiliation{Instituto de Física, Universidad Nacional Autónoma de México,
Apartado Postal 20-364, Ciudad de México, México.}

\date{Today}

\begin{abstract}
We investigate the transport properties of active particles undergoing a three-state run-and-tumble dynamics in one dimension, induced by non-reciprocal transition rates between self-propelling velocity states $\{-v, 0, +v\}$ that explicitly break microscopic reversibility. Departing from conventional reciprocal models, our formulation introduces a minimal yet rich framework for studying non-equilibrium transport driven by internal state asymmetries. Using kinetic Monte Carlo simulations and analytical methods, we characterize the particle’s transport properties across the transition-rates space. The model exhibits a variety of non-equilibrium behaviors, including ballistic transport, giant diffusion, and Gaussian or non-Gaussian transients, depending on the degree of asymmetry in the transition rates. We identify a manifold in transition-rate space where long-time diffusive behavior emerges despite the absence of microscopic reversibility. Exact expressions are obtained for the drift, effective diffusion coefficient, and moments of the position distribution. Our results establish how internal-state irreversibility governs macroscopic transport, providing a tractable framework to study non-equilibrium active motion beyond reciprocal dynamics.
\end{abstract}

\maketitle

\section{Introduction} 
Active motion emerges in particles that consume energy locally and transform it into  motion through complex self-propelling mechanisms that inherently operate far from equilibrium \cite{SchweitzerPhysRevLett1998,TailleurPRL2008,Lauga_RepProgPhys2009,RamaswamyAnnRevCondMattPhys2010,RomanczukEPJST2012,Bechinger2016}. A hallmark of active motion is its characteristic persistence, i.e., the tendency to maintain their direction of movement over a characteristic persistence time before reorienting. This leads to a wide range of transport phenomena not observed in equilibrium systems, such as enhanced diffusion \cite{KlumppPhysRevLett2005,HowsePhysRevLett2007}, clustering \cite{Buttinoni_2013,Bialke2015,Molodtsova2025}, and anomalous fluctuations \cite{Spinney_2025}.

One of the simplest recurred model of active motion is the so-called run-and-tumble, originally introduced to describe bacterial locomotion \cite{berg1993random,BarbaraFEMSMicrobioEco2003,ThevesBiophysJ2013,TaktikosPlos2014}. In its minimal form, this one-dimensional model considers particles that alternate stochastically between two velocity states, $\{+v, -v\}$, with exponential waiting times \cite{IoanaBena_2006,RossettoJStatMech2018}. Despite its simplicity, this binary-state model captures key features of active transport, including a crossover from ballistic motion at short times to diffusive behavior at long times \cite{Codling2008,Doussal2019}, with an effective diffusion coefficient that depends on the persistence rate.

Extensions to more than two velocity states have been recently explored to incorporate richer self-propulsion dynamics, particularly relevant, but still simple, are three-state models that include a resting state ($v = 0$) in addition to the two active states \cite{HahnPRE2020,BasuJPhysA2020,FrydelJStatMech2021,Fedotov_2022,Breoni2022}, which have led to nontrivial transport properties, as well as a better understanding of emergent transport regimes in active particles.

These models account for intermittent behavior observed in various biological and synthetic systems, where motile particles switch between movement and pauses \cite{Benichou2006,Wu-Zhang2025}. However, prior studies typically assume \emph{reciprocal transitions} between states---i.e., the rate of transition from state $n$ to $m$ equals that of the reverse process---thus preserving a form of microscopic reversibility in the internal dynamics \cite{YouPNAS2020}.

In this work, we depart from that assumption and consider a three-velocity-states run-and-tumble model in which the six possible transitions between the internal velocity states are \emph{non-reciprocal}, explicitly breaking detailed balance. This inherently non-equilibrium property drives the model beyond the standard non-equilibrium regime, resulting in qualitatively distinct behavior and allows us to explore the role of internal irreversibility to induce specific large-scale transport properties. We use of a combination of analytical techniques and kinetic Monte Carlo simulations, to show that such asymmetry gives rise to a broad spectrum of non-equilibrium phenomena including long-time ballistic transport, giant diffusion, and non-Gaussian displacement statistics.

We also elucidate a well-defined manifold in transition-rate space where long-time diffusive behavior is restored, despite the absence of microscopic reversibility. This reveals how macroscopic equilibrium-like behavior can emerge from fundamentally irreversible dynamics at the microscopic level. Our model thus serves as a minimal yet versatile framework for studying irreversible active transport, with potential implications for both biological systems and engineered active matter.

%%%%%%%%%%%%%%

The paper is organized as follows: in Sect.~\ref{SectII}  the run-and-tumble model of three internal motility states is presented. The marginal probability density $p(x;t)$ of finding an active particle at the position $x$ at time $t$, independent of the internal self-propelling state, is exactly obtained in Fourier-Laplace variables for the active process, the passive part of the motion described by thermal noise is incorporated through a convolution with the Gaussian propagator. Relevant quantities that characterized the transport properties are identified. In Sect.~\ref{Sect:Moments} the firsts moments of $p(x;t)$ are  computed and analyzed, and particular attention is given to transport properties deduced form the mean squared displacement and the kurtosis. Particular cases for three and four non-reciprocal transitions rates are analyzed in Sect.~\ref{Sect:ParticularCases}. We give our conclusions and final remarks in Sect.~\ref{Sect:Conclusions}.

%%%%%%%%%%%%%%%%%%%%%%%%%%%%%%%%%%%%%%%%%%%%%        SECTION II     

\section{\label{SectII}Three states run-and-tumble model}

We consider run-and-tumble particles moving in the infinite domain $(-\infty,\infty)$ in one dimension whose self-propulsion state takes one of the three velocity states  \cite{HahnPRE2020,BasuJPhysA2020,FrydelJStatMech2021,Fedotov_2022,Breoni2022}: $\{v_{+} = v,v_{0} =0,v_{-} =-v\}$ (with $v>0$ a constant, the self-propulsion speed), the self-propelling states $\pm v$ are the usual ones of run-and-tumble motion, while $v=0$ describes the state of tumbling for which the particle rests. Patterns of motion that resembles this one have been observed in \emph{Escherichia coli} bacteria, \emph{Caulobacter crescentus} and \emph{Chlamydomonas reinhartii} see Ref. \cite{Breoni2022} and references therein, or in the motion of  cytoskeletal motor proteins \cite{HafnerSciReps2016}. In addition, when under the effects of thermal noise, the inclusion of the self-propulsion state with vanishing velocity leads to an intermittent dynamics between active and Brownian motion. 

\begin{figure}[h]
\includegraphics[scale=0.23]{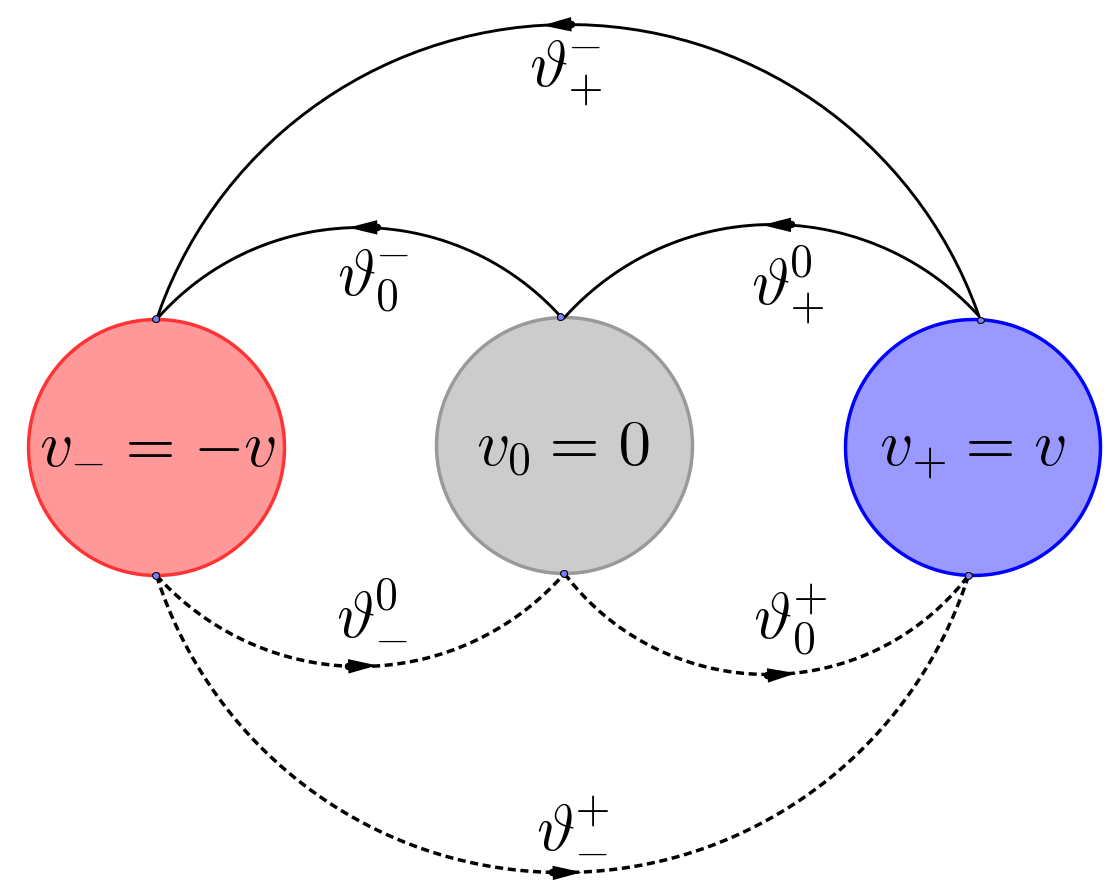}
\caption{The three velocity states $\{v_{-},v_{0},v_{+}\}$ are connected by the transition rates $\vartheta^{m}_{n}$, $n\neq m$, which are represented by solid and doted lines pointing out the direction of the transition and the reversed ones, respectively.}
\label{fig:Transition_rates}
\end{figure}

The transition rates $\vartheta_n^m$ from the self-propulsion state $n$ to state $m$, where $n,m\in \mathbb{S}=\{+,0,-\}$, are time-space independent and non-reciprocal, meaning that $\vartheta_-^+ \neq\vartheta_+^-$ for the transitions between the states $\{v_-,v_+\}$;  $\vartheta_-^0\neq\vartheta_0^-$ for the transitions between states $\{v_-,v_0\}$; and  $\vartheta_0^+\neq\vartheta_+^0$ for the transitions between the states $\{v_0,v_+\}$ (Fig.~\ref{fig:Transition_rates} depicts such a situation), thus breaking microscopic reversibility and extending the situation studied by Hahn et al. \cite{HahnPRE2020}, where only reciprocal transitions between states were considered. We consider that once the particle internal state is in one velocity state, it transits only to one of the other two, i.e., transitions to the same state are not possible.

\subsection{The stochastic process under study}
The stochastic dynamics of the particle position, $x(t)$, is determined from the stochastic differential equation 
\begin{equation}\label{Langevin}
\frac{d}{dt}x(t)=v_\textrm{sp}(t)+\xi(t),
\end{equation}
where the self-propulsion velocity $v_\textrm{sp}(t)$, is a trichotomous stochastic process whose time evolution describes the transition among the states $\{-v,0,+v\}$, with the known transition rates $\{\vartheta_n^m\}$, $n,m\in\mathbb{S}$. The particle motion is also affected by a thermally fluctuating process $\xi(t)$, modeled by Gaussian white noise, i.e., $\langle\xi(t)\rangle=0$ and $\langle\xi(t) \xi(s)\rangle = 2D\delta(t-s)$, $D=k_{B}T/\gamma$ being the diffusion coefficient and $\gamma$ the friction coefficient that emerges from the interaction between the particle and the bath. Our statistical analysis is carried out through the ensembles of trajectories obtained by numerically solving equation \eqref{Langevin} using the time-independent version of the widely used Kinetic Monte Carlo (KMC) method \cite{BortzJCompPhys1975}.

\subsection{\label{subsect:FPE} The associated Fokker-Planck equation to the stochastic process under study}
The Fokker-Planck equation associated to the stochastic differential equation \eqref{Langevin}, for the joint probability density $P(x,v_{n};t)$ of a run-and-tumble particle moving in one dimension, located at $x\in(-\infty,\infty)$ and moving with velocity $v_{n}$ with $n\in\mathbb{S}$ at time $t$ is given by:
\begin{multline}\label{FokkerPlanckEq}
 \frac{\partial}{\partial t}P(x,v_{n};t)+v_{n}\frac{\partial}{\partial x}P(x,v_{n};t)=D\frac{\partial^{2}}{\partial x^{2}}P(x,v_{n};t)\\
 +\sum_{m\in \mathbb{S} }\left[%\mathbb{T}
 \vartheta_{m}^{n} P(x,v_{m};t) -%\mathbb{T}
 \vartheta_{n}^{m} P(x,v_{n};t)\right],
\end{multline}
where $%\mathbb{T}
\vartheta^{m}_{n}$ denotes the entries of the transition matrix $\boldsymbol{T}$,
\begin{equation}\label{t_matrix}
\boldsymbol{T}=\begin{pmatrix}
0 & \vartheta_{-}^{0} & \vartheta_{-}^{+}\\
\vartheta_{0}^{-} & 0 & \vartheta_{0}^{+}\\
\vartheta_{+}^{-} & \vartheta_{+}^{0} & 0
\end{pmatrix}
\end{equation}
that embodies the time-independent transition rates $\vartheta_{n}^{m} \ge 0$ between distinct particle velocity states $n$ and $m$ $\in\mathbb{S}$ as shown in Fig.\eqref{fig:Transition_rates}. 

\subsection{The marginal probability distribution of the self-propulsion states $p_n(t)$}\label{Sect:StationarySelfPropulsionSates}
After integrating Eq.~\eqref{FokkerPlanckEq} over the whole spatial domain, we obtain the corresponding equation for the marginal distribution $p_n(t)=\int_{-\infty}^\infty dx\,  p_n(x;t)=\int_{-\infty}^\infty dx\, P(x,v_n;t)$
\begin{equation}
\frac{d}{dt}p_n(t)=\sum_{m\in \mathbb{S} }\left[%\mathbb{T}
\vartheta_{m}^{n} p_m(t) -%\mathbb{T}
\vartheta_{n}^{m} p_n(t)\right],
\end{equation}
which acquires the stationary state as $t\rightarrow\infty$
\begin{subequations}
\label{Stat}
\begin{align}
p_-^\text{st}&=\frac{\lambda_-}{\lambda}=\frac{\vartheta_0^-\vartheta_+^0+\vartheta_0^-\vartheta_+^-+\vartheta_0^+\vartheta_+^-}{\lambda},\\
p_0^\text{st}&=\frac{\lambda_0}{\lambda}=\frac{\vartheta_-^0\vartheta_+^0+\vartheta_-^+\vartheta_+^0+\vartheta_-^0\vartheta_+^-}{\lambda},\\
p_+^\text{st}&=\frac{\lambda_+}{\lambda}=\frac{\vartheta_0^+\vartheta_-^0+\vartheta_0^-\vartheta_-^++\vartheta_0^+\vartheta_-^+}{\lambda},
\end{align}
\end{subequations}
with 
\begin{equation}
\label{Lambda}
\lambda\equiv\lambda_-+\lambda_0+\lambda_+,
\end{equation}
the normalization factor or \emph{partition function} of the statistical weights $\{\lambda_-,\lambda_0, \lambda_+\}$. 

It can be shown that the time dependence of the solutions $p_n(t)$ involves the time scale $\Theta^{-1}$, where
\begin{align}\label{TimeScale}
\Theta \coloneqq \sum_{n, m}\vartheta_m^n,
\end{align}
this defines the relaxation time towards stationary distribution of the self-propulsion states, and thus is identified with the time scale. This characteristic time defines the characteristic length scale $\ell \coloneqq v\Theta^{-1}$. 

The dimensionless form of Eq.~\eqref{TimeScale} corresponds to the equation of a 5-\emph{simplex} (\emph{hexateron}, $\mathbb{H}$) that defines the set of six-tuples $\{\widetilde{\vartheta}_-^+,\widetilde{\vartheta}_-^0,\widetilde{\vartheta}_0^-,\widetilde{\vartheta}_0^+,\widetilde{\vartheta}_+^0,\widetilde{\vartheta}_+^-\}$ that satisfy $\sum_{n\neq m}\widetilde{\vartheta}_m^n=1$, $0\le\widetilde{\vartheta}_m^n \le1$, $\widetilde{\vartheta}_m^n\coloneqq\vartheta_m^n/\Theta$ being the dimensionless transition rates. This allows a systematic analysis of the particle's transport properties induced by the dimensionless transition rates in terms of the six-dimensional parameter space $\widetilde{\vartheta}_m^n$, since a physically-meaningful decomposition of the 5-simplex into smaller simplices is possible. 
Explicitly, there are six cases for which one rate can be chosen to vanish (one of the six edges in Fig.~\ref{fig:Transition_rates} can be deleted), these correspond to the six 4-simplices. There are
15 ways for which two rates can be chosen to vanish (two edges in Fig.~\ref{fig:Transition_rates} can be deleted), each corresponding to one of the 15 3-simplices (tetrahedral faces). There 20 ways to chose three rates in the system to vanish (number of ways of deleting three edges in Fig.~\ref{fig:Transition_rates}) these are the 20 2-simplices or triangular faces of the 5-simplex. Likewise, the are 15 cases of choosing 4 transitions rates to vanish; and 6 ways of choosing 5 transitions to vanish. Although the parameter space has a simple shape (is a six-dimensional hyperplane), it is still too large to allow for an exhaustive analysis. We therefore use a set $\mathbb{T}$ of $10^6$ uniformly and randomly sampled six-tuples of the variables $\widetilde{\vartheta}_m^n$ [this corresponds to the uniform generation of dimensionless random matrices $\boldsymbol{T}$ of the form \eqref{t_matrix}], to assess the distribution of characteristic quantities, as will be discussed below. 

As will be shown afterwards, the long-time properties of the particle's transport explicitly depend on the the stationary probability distribution of the self-propelling states given by Eqs.~\eqref{Stat}. Here we would like to point out that ballistic transport in the long-time regime will be observed for the majority of the transitions rates values, and importantly, that diffusive transport is possible in a variety of cases even when non-reciprocal transitions rates are chosen, and therefore it can be specified by properly choosing the values of the transition rates, as will be clear in the following sections.

\begin{figure}
\includegraphics[width=0.95\columnwidth]{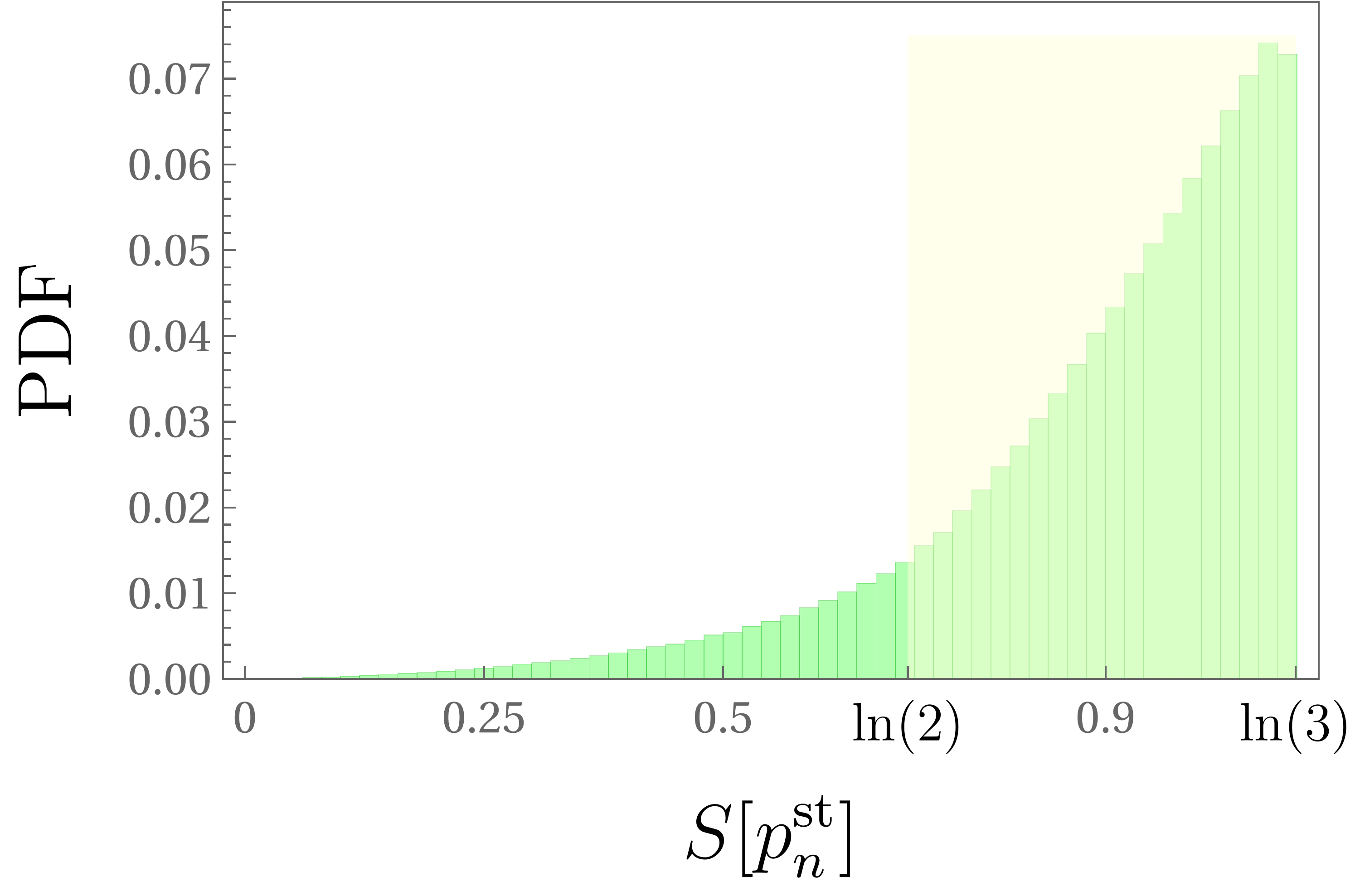}
\caption{Distribution of the information entropy $S[p_n^\textrm{st}]$ obtained from a uniform sample of $10^6$ six-tuples $\widetilde{\vartheta}_n^m$ in the 5-simplex (see text).}
\label{fig:Entropy}
\end{figure}

The entropy of information of the stationary distribution of the self-propulsion states: $S[p_n^\textrm{st}]\coloneqq-\sum_{n\in\mathbb{S}}p_n^\textrm{st}\ln p_n^\textrm{st}$ provides some general guidance about the transport behavior---either diffusive or ballistic---of the particles. The distribution $p_{n}^\textrm{eq}$ that maximizes $S[p_n^\textrm{st}]$, the \emph{equilibrium} probability distribution is given by $p_-^\textrm{eq}$, $p_+^\textrm{eq}$, $p_0^\textrm{eq}=\frac{1}{3}$, which can be shown is in direct correspondence with the detailed-reciprocity case, i.e., $\widetilde{\vartheta}_n^m=\frac{1}{6}$ that coincides with the symmetry center of the 5-simplex (barycenter). This point of symmetry induces normal diffusion as the characteristic long-time regime transport, since moving to the right or to the left is statistically symmetric and thus no bias about the direction of motion is expected. In Fig.~\ref{fig:Entropy} the distribution of values of $S[p_n^\textrm{st}]$ is shown from a sample of $10^6$ six-tuples $\widetilde{\vartheta}_n^m$ uniformly chosen in the 5-simplex. It is shown a maximum of six-tuples around the equilibrium value of entropy $\ln3$.

Deviations from the detailed-reciprocity condition, would induce either a long-time ballistic transport or a long-time rest state (particle relaxes to the zero velocity state). The first is characterized by a
drift velocity proportional to $p_+^\text{st}-p_-^\text{st}\neq0$ (or equivalently to $\lambda_--\lambda_+\neq0$) that biases the direction of motion. It is well expected, and we will show, that the condition of vanishing of the drift velocity, i.e. $p_+^\text{st}=p_-^\text{st}$ is sufficient to lead to normal diffusion in the long-time regime. Such a condition allows the appearance of normal diffusion not only for maximum entropy, $\ln(3)$ which indicates the equilibrium distribution of the self-propulsion states, but for a wide range of values of it, $S[p_n^{\textrm{st}}]\in[\ln(2),\ln(3)]$, that corresponds to non-equilibrium stationary distribution  of the self-propelling states.
The loss of micro-reversibility induced by the non reciprocal relations among the transitions rates between self-propulsion states. In the following we carry out an analysis of the effects of the departure from complete reciprocity.

%%%%%%%%%%%%%
\subsection{The marginal probability densities of the particle positions $P(x;t)$ and $p(x;t)$}
The additive nature of the right-hand side of Eq.~\eqref{Langevin} suggests to write $P(x,v_n;t)$ as the convolution of the Gaussian propagator $G_{D}(x;t)=\exp\{-x^{2}/4Dt\}/\sqrt{4\pi Dt}$ [solution of the diffusion equation $\partial_{t}G(x;t)=D\partial^{2}_{xx}G(x;t)$ with initial conditions $G_{D}(x;0)=\delta(x)$], with the active probability density $p_n(x;t)$, i.e., 
\begin{equation}\label{Convolution}
P(x,v_{n};t)=\int_{-\infty}^{\infty}dx^{\prime}G_{D}(x-x^{\prime};t)p_{n}(x^{\prime};t),
\end{equation}
with $n\in\mathbb{S}$ and $p_n(x;t)$ satisfying
\begin{widetext}
\begin{subequations}\label{3stat-masteq}
\begin{align}
\frac{\partial}{\partial t}p_-(x;t) - v\frac{\partial}{\partial x} p_{-}(x;t)&=\vartheta_+^-\, p_+(x;t)+\vartheta_0^-\, p_0(x;t)-(\vartheta_{-}^{0}+\vartheta_-^+)\, p_-(x;t),\\
\frac{\partial}{\partial t}p_0(x;t)&=\vartheta^{0}_{-}\, p_-(x;t)+\vartheta^{0}_{+}\, p_+(x;t)-(\vartheta_{0}^{-}+\vartheta_{0}^{+})\,  p_0(x;t),\\
\frac{\partial}{\partial t}p_+(x;t)+v\frac{\partial}{\partial x}p_+(x;t)&=\vartheta_{-}^{+}\, p_-(x;t)+\vartheta_{0}^{+}\, p_0(x;t)-(\vartheta_{+}^{-}+\vartheta_{+}^{0})\,  p_+(x;t).
\end{align}
\end{subequations}
\end{widetext}
Thus, we focus our analysis on the Eqs.~\eqref{3stat-masteq}. Notice that after adding them, the continuity equation appears:
\begin{subequations}
\begin{equation}
\pdt p(x;t)+\pdx j(x;t)=0,
\end{equation}
where $p(x;t)=p_-(x;t)+p_0(x;t)+p_+(x;t)$
and the probability current $j(x;t)$ is given by
\begin{equation}
    j(x;t)\equiv v\bigl(p_+(x;t)-p_-(x;t)\bigr).
\end{equation}
\end{subequations}

An analysis of the solutions of the linear system of partial differential equations is rather cumbersome, thus we resort to an analysis in Fourier-Laplace domain, where the description in space-time variables $x$-$t$ is transformed to the description in terms of the variables $k$-$\epsilon$  through the transformation $\tilde{f}(k;\epsilon)=\mathcal{L}\{\hat{f}(k;t)\}=\mathcal{L}\{\mathcal{F}\{f(x;t)\}\}$ 
where $\mathcal{L}\{g(t)\}=\int_0^\infty dt\, e^{-\epsilon t}g(t)$ and $\mathcal{F}\{h(x)\}=\int_{-\infty}^{\infty}dx\, e^{-ikx} h(x)$. The value of such analysis will be evident in the following sections, where the intermediate scattering function that corresponds to the Fourier transform of the probability density of finding a particle at the position $x$ at time $t$, independent of the self-propulsion state, is analyzed. After taking the Fourier-Laplace transform to the set of equations \eqref{3stat-masteq}, we get the following linear system of algebraic equations:

\begin{subequations}\label{3statesRTP-FL}
\begin{align}
(\epsilon-\text{i} kv+\vartheta_{-}^{0}+\vartheta_{-}^{+})\widetilde{p}_--\vartheta_0^-\widetilde{p}_0-\vartheta_+^-\widetilde{p}_+& =\hat{p}^{(0)}_-,\\
-\vartheta^{0}_{-}\widetilde{p}_-+(\epsilon+\vartheta_{0}^{-}+\vartheta_{0}^{+})\widetilde{p}_0-\vartheta^{0}_{+}\widetilde{p}_+& =\hat{p}_0^{(0)},\\
-\vartheta_{-}^{+}\widetilde{p}_{-} - \vartheta_{0}^{+}\widetilde{p}_{0} + (\epsilon+\text{i} kv+\vartheta_{+}^{0}+\vartheta_{+}^{-})\widetilde{p}_{+}& = \hat{p}_{+}^{(0)},
\end{align}
\end{subequations}
where we have omitted the arguments $(k;\epsilon)$ for the sake of abbreviation and $\hat{p}_{n}^{(0)}=\hat{p}_{n}^{(0)}(k)=\hat{p}_{n}(k;t=0)$, $(n\in\mathbb{S})$ denote the initial conditions. 

\subsection{The marginal probability density $p(x;t)$ and the intermediate scattering function $\hat{p}(k;t)$}
We are interested in the transport properties of such a model, given by the first moments of the spatial distribution $p(x;t)=p_-(x;t)+p_0(x;t)+p_+(x;t)$ that corresponds to the probability density of finding a particle in position $x$ independent of the state of self-propulsion. After solving  Eqs.~\eqref{3statesRTP-FL} for $\widetilde{p}_+(k;\epsilon)$, $\widetilde{p}_0(k;\epsilon)$ and $\widetilde{p}_-(k;\epsilon)$, and after a carefully rearrangement of the terms to compute $\widetilde{p}(k;\epsilon)$, we get the simple form
\begin{multline}\label{MarginalP-FL}
\widetilde{p}(k;\epsilon)=\frac{\hat{p}^{(0)}(k)}{\epsilon+\widetilde{F}(k;\epsilon)\bigl[\text{i} kv_\text{drift}+ k^2 D_\text{act}\bigr]}+\\
\frac{\Theta^{-2}\widetilde{F}(k;\epsilon)\bigl[(\epsilon+\Theta)\partial_t \hat{p}^{(0)}(k)+\partial_{tt}\hat{p}^{(0)}(k)\bigr]}{\epsilon+\widetilde{F}(k;\epsilon)\bigl[\text{i} kv_\text{drift}+ k^2 D_\text{act}\bigr]},
\end{multline}
where the dimensionless $\widetilde{F}(k;\epsilon)
$ is defined by
\begin{equation}\label{Memory}
\widetilde{F}(k;\epsilon)\coloneqq\frac{\Theta^2}{\bigl(\epsilon+\frac{\Theta}{2}\bigr)^2+\Delta^2+k^2v^2+\Theta\textrm{i}kv_\textrm{act}}.
\end{equation}
\begin{figure}
\includegraphics[width=0.9\columnwidth]{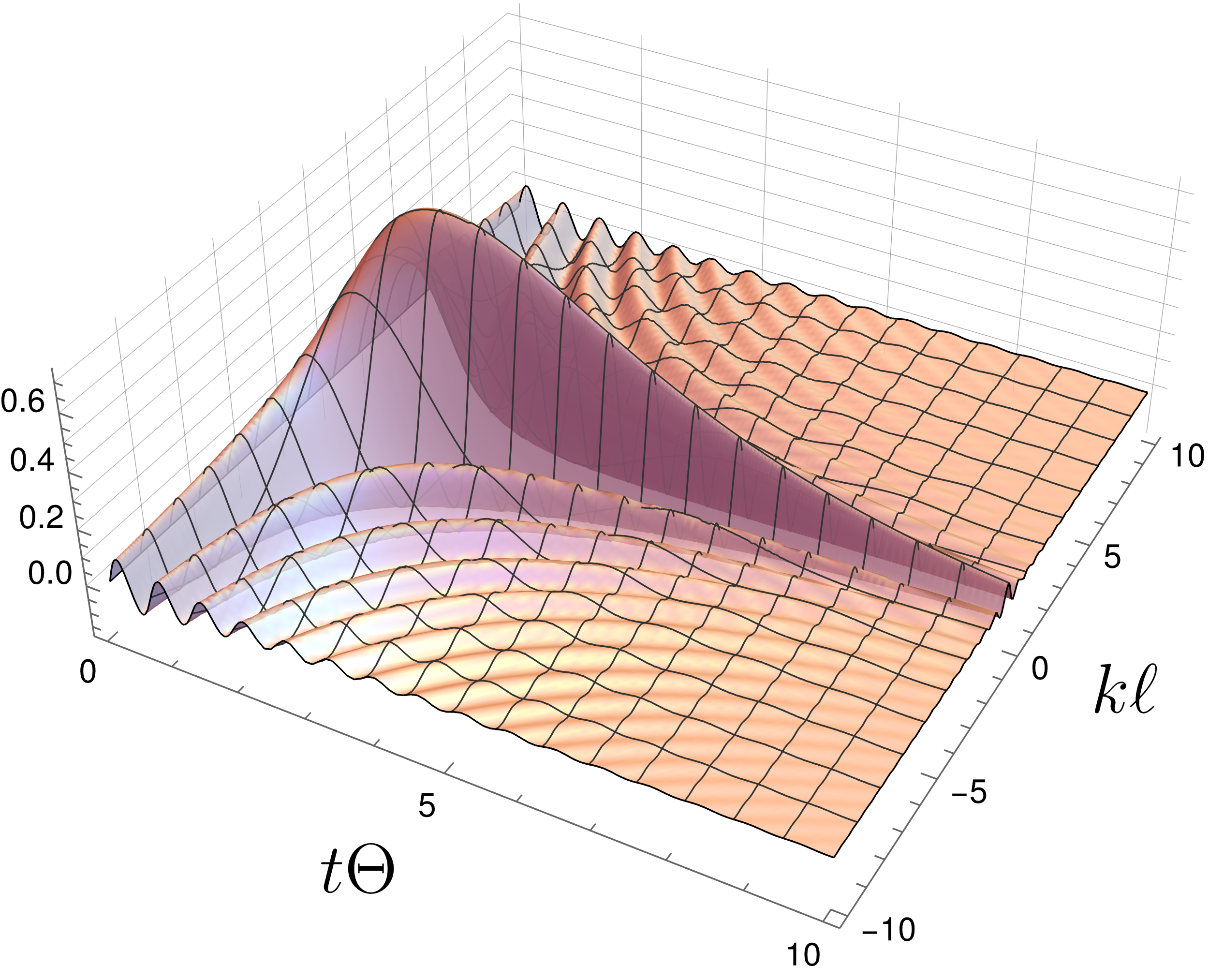}\\
\includegraphics[width=0.9\columnwidth]{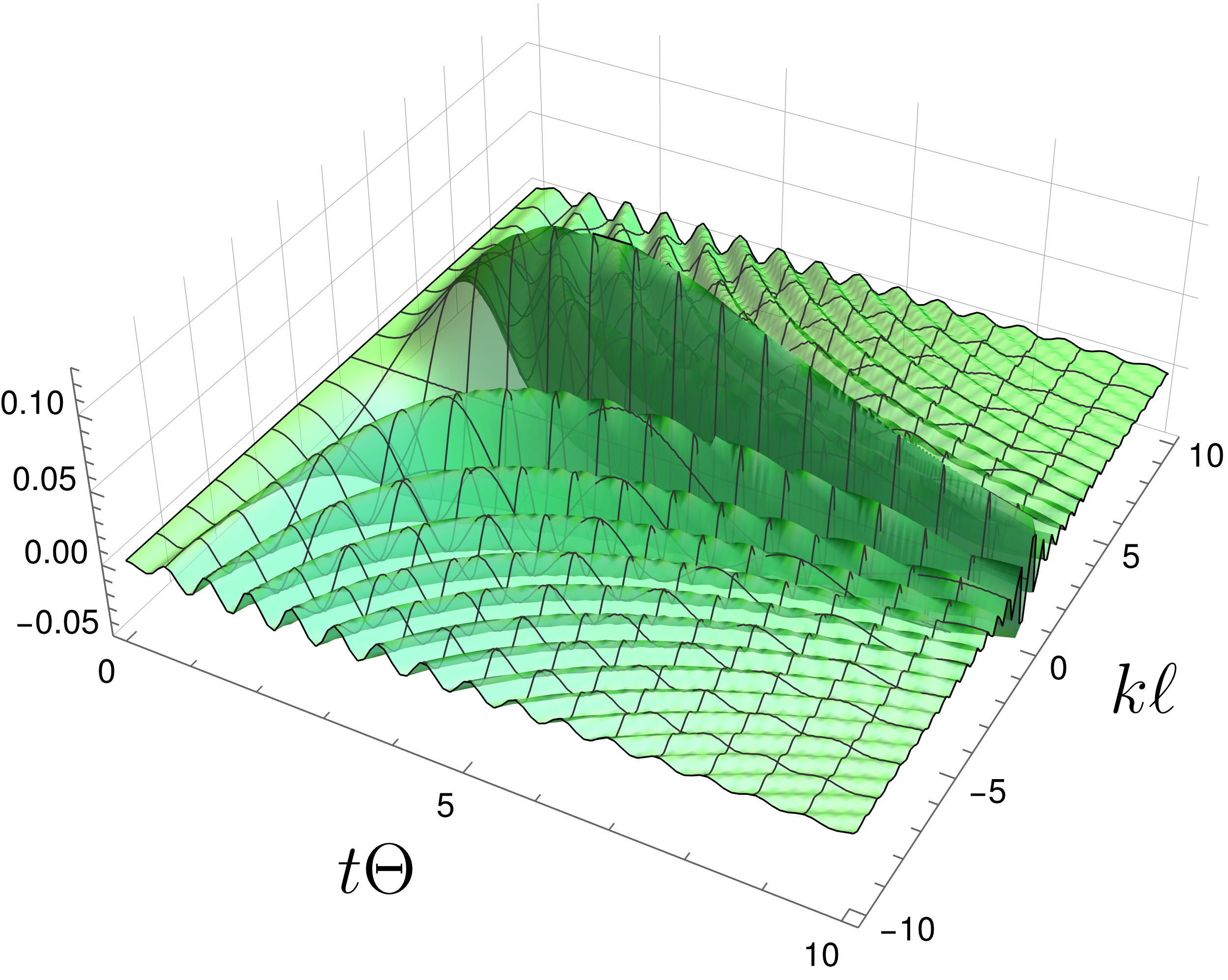}
\caption{Typical dependence on time ($t\Theta$) and wavenumber ($k\ell$) of the real (top) and imaginary (bottom) parts of the memory function $\hat{F}(k;t)$ for a typical six-tuple uniformly chosen at random.}
\label{fig:Memory}
\end{figure}
$\hat{F}(k;t)$ is shown in Fig.\ref{fig:Memory} after numerically inverting the Laplace transform for a typical six-tuple of transition rates. 

In Eq.~\eqref{MarginalP-FL}, $\hat{p}^{(0)}(k)$ corresponds to the Fourier transform of the initial data $p(x;0)=p_-(x;0)+p_0(x;0)+p_+(x;0)$, while 
\begin{equation*}
\partial_t\hat{p}^{(0)}(k)\coloneqq\frac{\partial}{\partial t}\hat{p}(k;t)\biggl\vert_{t=0},\, \partial_{tt}\hat{p}^{(0)}(k)\coloneqq\frac{\partial^2}{\partial t^2}\hat{p}(k;t)\biggl\vert_{t=0},
\end{equation*}
can be computed explicitly in terms of the initial information $p_-(x,0)$, $p_0(x,0)$, $p_+(x,0)$, by use of the continuity equation (see the appendix \ref{app:InitialConditions}). 

\begin{figure}
\includegraphics[width=0.9\columnwidth]{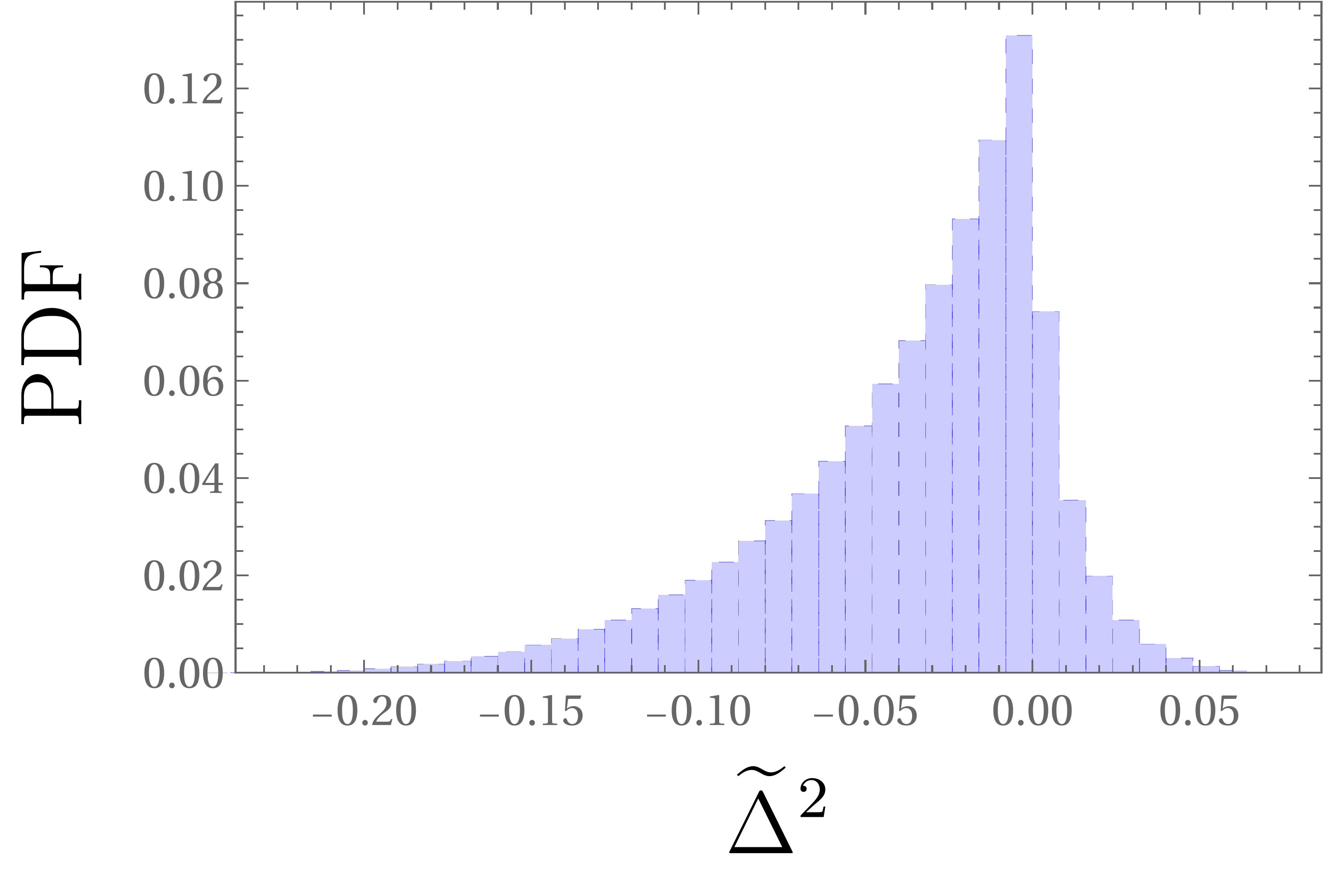}
\caption{The distribution of values of the dimensionless $\widetilde{\Delta}^2$ is shown from a uniform sample ($\mathbb{T}$) of dimensionless transition rates in the 5-simplex $\mathbb{H}$.}
\label{fig:Delta2}
\end{figure}

Different characteristic quantities, defined in terms of the transitions rates, emerge in the solution \eqref{MarginalP-FL}. \begin{subequations}\label{Parameters}
$\Theta^{-1}$ is the natural characteristic time scale for the system dynamics, as was indicated in the last section \ref{subsect:FPE}; 
while $\lambda$ is the partition function of the stationary distribution of the self-propulsion state, independent of the particle position, as given in Eqs.~\eqref{TimeScale} and \eqref{Lambda}, respectively. $D_\text{act}$ is a diffusion coefficient that emerges from the randomization process of the particle state of motion given by 
\begin{align}\label{Dact}
D_\textrm{act}\coloneqq\dfrac{v^2}{\Theta^2} (\vartheta_{0}^{-}+\vartheta_{0}^{+}).
\end{align}
%%%%
\begin{figure}
\includegraphics[width=0.9\columnwidth]{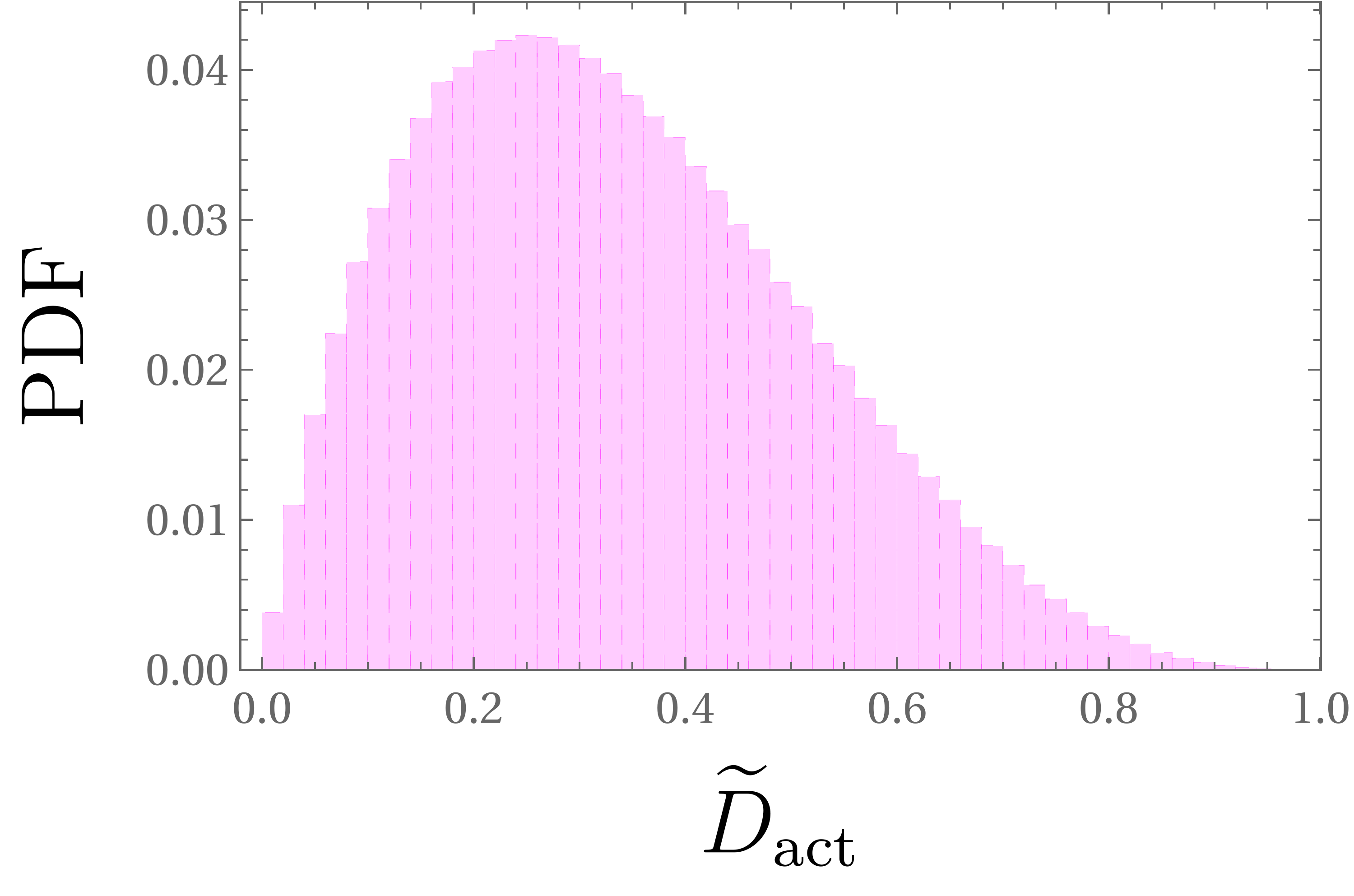}
\caption{Distribution of values of the dimensionless active diffusion $\widetilde{D}_\textrm{act}$ obtained from a uniform sample of the dimensionless transition rates $\widetilde{\vartheta}_n^m$, $\mathbb{T}$.
}
\label{fig:Dact}
\end{figure}
The dependence only on $\vartheta_0^+$ and $\vartheta_0^-$ indicates that diffusion always contributes to the transport properties except when both rates vanish, in such a case the stationary state of self-propelling velocities relaxes to the rest state $p_\pm^\textrm{st}=0$, $p_0^\textrm{st}=1$. In Fig.~\ref{fig:Dact} we show the distribution of values of $\widetilde{D}_\textrm{act}\coloneqq D_\textrm{act}/(v^2\Theta^{-1})=\widetilde{\vartheta}_0^-+\widetilde{\vartheta}_0^+$ from a uniform sample of six-tuples $\widetilde{\vartheta}_n^m$, $\mathbb{T}$ in the 5-simplex, a maximum 0.042305 is observed at $\widetilde{D}_\textrm{act}\approx 0.25$ with a mean value $0.33352\pm0.1783$. 

In addition two characteristic velocities appear, one is denoted as the \emph{active velocity}
\begin{equation}\label{vact}
 v_\textrm{act}\coloneqq\dfrac{v}{\Theta}\bigl(\vartheta_{-}^{0}+ \vartheta_{-}^{+}-\vartheta_{+}^{0} -\vartheta_{+}^{-}\bigr)
\end{equation}
that sizes the difference between the total transition rate from velocity state $-v$,  $\vartheta_{-}^{0}+ \vartheta_{-}^{+}$, from the total transition rate from velocity state $+v$,  $\vartheta_{-}^{0}+ \vartheta_{-}^{+}$. The other characteristic velocity corresponds to the \emph{drift velocity}
\begin{align}\label{Eq:vdrift}
 v_\text{drift}&\coloneqq v\frac{\lambda_+-\lambda_-}{\Theta^2}=v\frac{\lambda}{\Theta^2}\bigl(p_+^\text{st}-p_-^\text{st}\bigr)\\ 
 &=\frac{v}{\Theta^2}\Bigl[\vartheta_0^+\vartheta_-^0+\vartheta_0^-\vartheta_-^++\vartheta_0^+\vartheta_-^+\label{Vdrift}\\
 &\qquad\qquad\qquad
 -\vartheta_0^-\vartheta_+^0-\vartheta_0^-\vartheta_+^--\vartheta_0^+\vartheta_+^-
 \Bigr],\nonumber
\end{align}
which emerges whenever $p_+^\text{st}\neq p_-^\text{st}$. As will be discussed immediately after this section, the manifold in the space of transition rates defined by $p_+^\text{st}=p_-^\text{st}$  corresponds to the locus that leads to long-time regime diffusive behavior as will be shown in Sect.~\ref{subsect:LongShortRegimes}. The distribution of the values of these velocities obtained in the sample $\mathbb{T}$ is shown in Fig.~\ref{fig:Characteristic-vels}. In both scenarios, the likelihood of  producing a vanishing velocity is maximum in the sample set $\mathbb{T}$, however, while $|v_\textrm{act}|$ may acquire any value in $[0,1]$, in contrast $|v_\textrm{drift}|$ can only acquire speed values below $0.25\, v$.
%%%%%%%
\begin{figure}
\includegraphics[width=0.9\columnwidth]{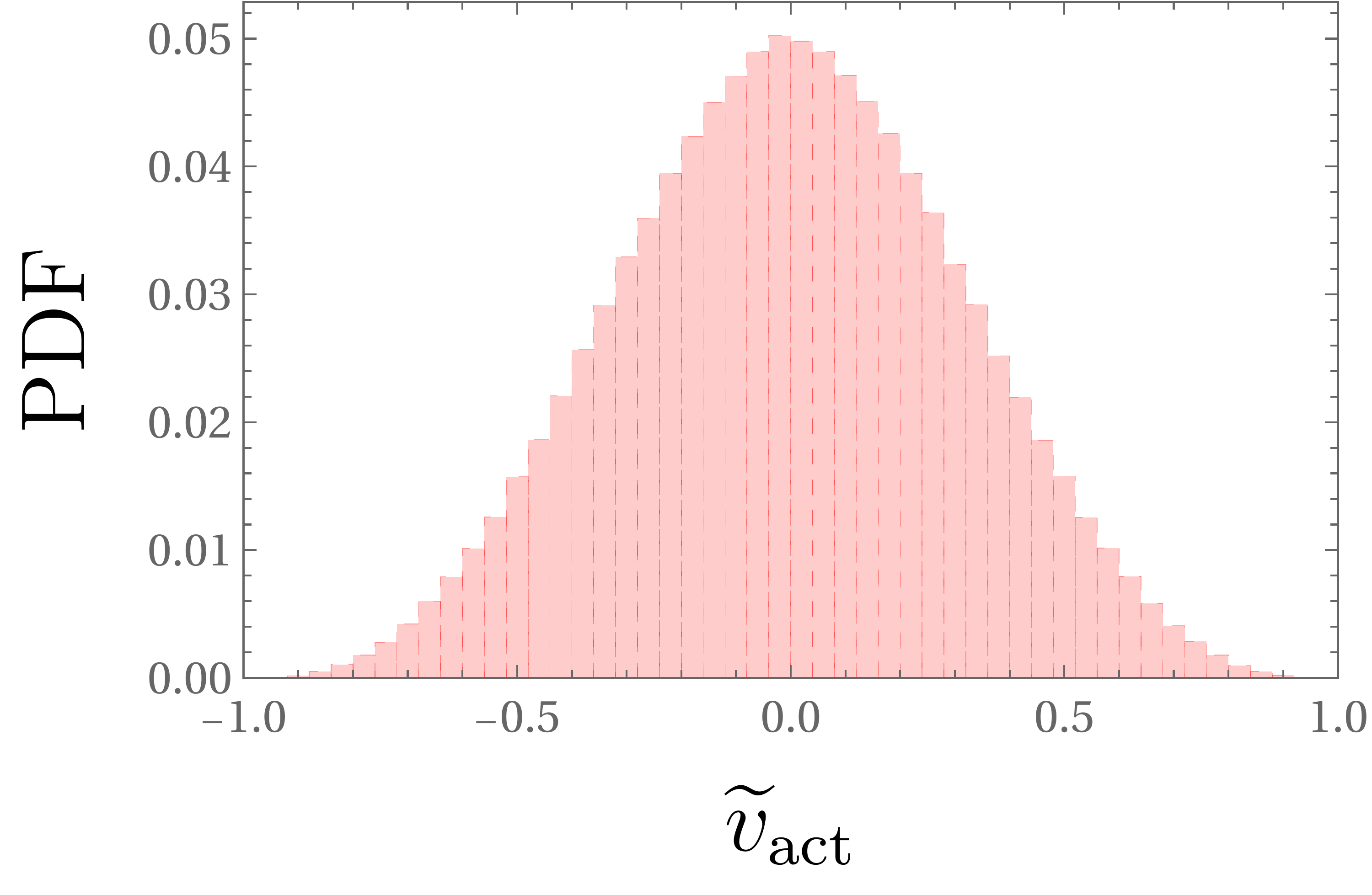}\\
\includegraphics[width=0.9\columnwidth]{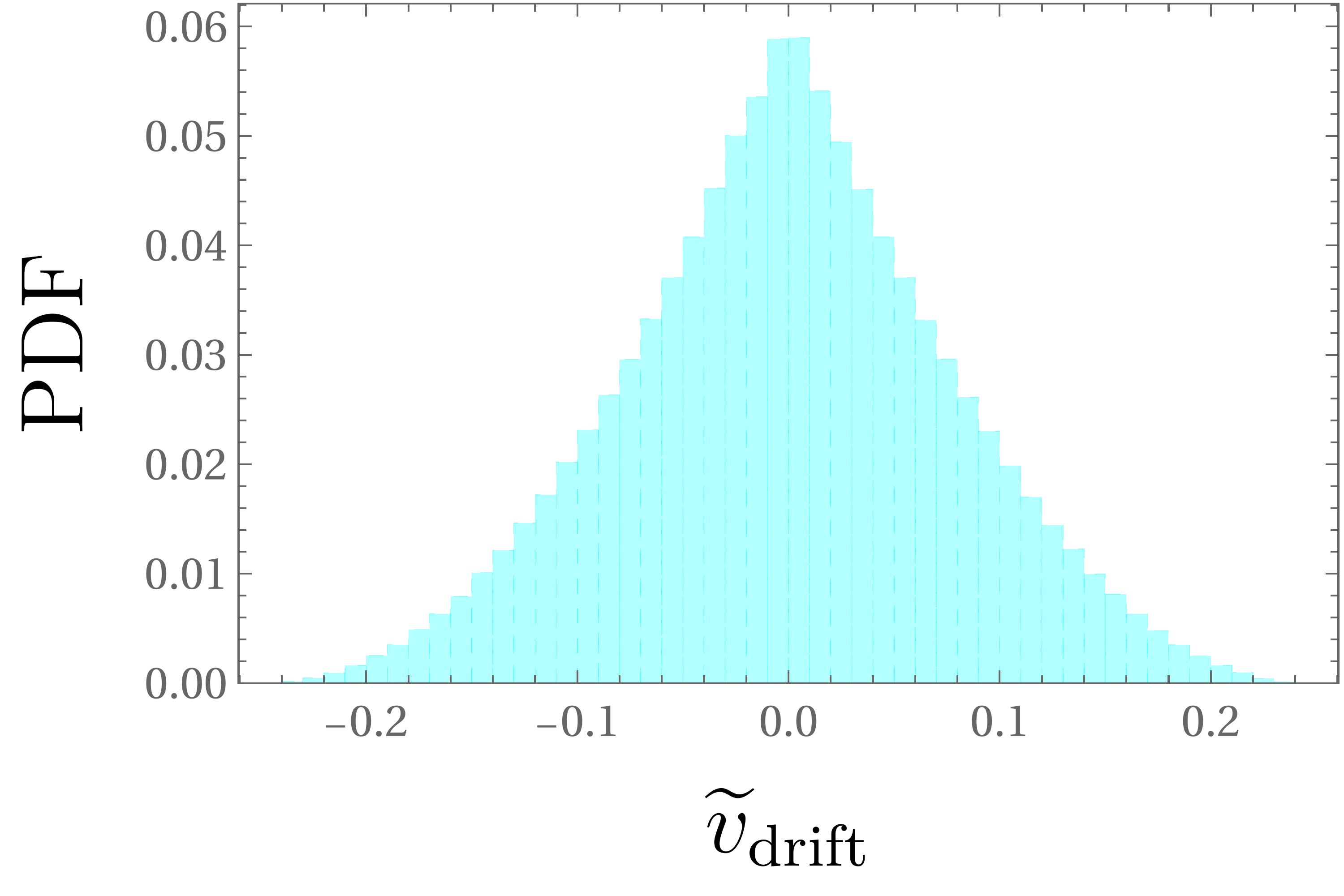}
\caption{Distribution of the characteristic (dimensionless) velocities $\widetilde{v}_\textrm{act}$ (top panel) and $\widetilde{v}_\textrm{drift}$ (bottom panel) in the set of uniformly sampled six-tuples $\mathbb{T}$. Both PDF's are symmetric around zero velocity where they are maximum, clearly indicating the most likely values.} 
\label{fig:Characteristic-vels}
\end{figure}

Finally
\begin{equation}\label{Delta2}
\Delta^2\coloneqq\lambda-\frac{\Theta^2}{4}   
\end{equation}
\end{subequations}
is a parameter that determines the specific system dynamics whether $\lambda>\frac{\Theta^2}{4}$ ($\Delta$ is real), $\lambda<\frac{\Theta^2}{4}$ ($\Delta$ is imaginary), or $\lambda=\frac{\Theta^2}{4}$ ($\Delta$ vanishes). This quantity has a relevant role in the transport properties as is discussed afterwards in this paper. It can be shown that in the case of the fully connected case (all $\vartheta_n^m\neq0$) the maximum value of $\Delta^2$ is zero for the fully symmetric case $\vartheta_n^m=\vartheta$; any other choice of $\vartheta_n^m$ case leads to $\Delta^2<0$. However, the range of values of $\Delta^2$ depends strongly on the non-vanishing transitions rates  as is shown later on in this paper. In Fig.~\ref{fig:Delta2} a distribution of values of the dimensionless $\widetilde{\Delta}^2\coloneqq\Delta^2/\Theta^2$ is shown for the uniform sample of $10^6$ six-tuples in the space of the dimensionless transition rates. Notice the negative values of $\Delta^2$ are the typical cases and a maximum at the value $\widetilde{\Delta}^2=0$ is conspicuous.

Initial information is propagated in time and space in different manners. The first term in the right-hand side of \eqref{MarginalP-FL} gives the evolution of $\hat{p}^{(0)}(k)$ through the propagator 
\begin{equation}
\widetilde{\mathcal{G}}(k;\epsilon)=\Bigl\{\epsilon+\widetilde{F}(k;\epsilon)\bigl[\text{i} kv_\text{drift}+ k^2 D_\text{act}\bigr]\Bigr\}^{-1},
\end{equation}
this satisfies a non-local in time and space diffusion-advection equation
\begin{equation}
\label{DA-Propagator}
\frac{\partial}{\partial t}\mathcal{G}(x;t)+v_\textrm{drift}\biggl(F**\frac{\partial}{\partial x}\mathcal{G}\biggr)=D_\textrm{act}\biggl(F**\frac{\partial^2}{\partial x^2}\mathcal{G}\biggr) 
\end{equation}
with initial condition $\mathcal{G}(x,0)=\delta(x-x^\prime)$ and $F(x;t)$ the memory function given by the inverse Fourier-Laplace transform of $\widetilde{F}(k;\epsilon)$ given in \eqref{Memory} and $(f**g)=\int_{-\infty}^\infty dx^\prime\int_0^t f(x-x^\prime;t-s)g(x^\prime;s)$ denotes the double convolution with respect to position and time variables. The second term in the right-hand side of \eqref{MarginalP-FL}, gives the propagation of the initial information $(\Theta+\epsilon)\partial_tp^{(0)}(k)+\partial_{tt}p^{(0)}(k)$ carried out by $\Theta^{-2}(\mathcal{G}**F)$.

The meaning of the solution \eqref{MarginalP-FL} can be elucidated by taking the inverse Fourier and Laplace transformations, that leads to the third-order partial differential equation 
\begin{widetext}
\begin{equation}\label{ThirdOrder-PDE}
       \dfrac{1}{\Theta^2}\dfrac{\partial}{\partial t}\biggl[\dfrac{\partial^2}{\partial t^2}+\Theta\dfrac{\partial}{\partial t}+\lambda+\Theta v_\textrm{act}\frac{\partial}{\partial x}-v^2\dfrac{\partial^2}{\partial x^2}\biggr]p(x;t)+ v_\text{drift}\dfrac{\partial}{\partial x}p(x;t) =D_\text{act}\dfrac{\partial^2}{\partial x^2}p(x;t),
\end{equation}
\end{widetext}
with initial condition $p^{(0)}(x)$, $\partial_t p^{(0)}(x)$ and $\partial_{tt}p^{(0)}(x)$ as already indicated at the beginning of this section. We point out in the subsequent section \ref{subsect:LongShortRegimes} that the third-order derivative in time that appears in Eq.~\eqref{ThirdOrder-PDE}, differs drastically from the wave-like (or ballistic) propagation observed in the short-time regime of the two-states run-and-tumble model \cite{BalakrishnanPhysicaA1988a}.

%%%%
\subsection{The long- and short-time regimes}
\label{subsect:LongShortRegimes}
\paragraph{Long-time and large-distance regime.-}
In the long-time ($\epsilon\ll\Theta$) and large-distance ($k\rightarrow0$) regimes, the memory function is local in time and space and therefore it can be approximated by  
\begin{align}
&\widetilde{F}(k;\epsilon)
\underset{\substack{\epsilon\rightarrow0 \\ k\rightarrow0}}{\rightarrow}
\frac{\Theta^2}{\lambda},
\end{align}
while for the initial data we have
\begin{subequations}
\begin{align}
&\hat{p}^{(0)}(k)\stackrel[k\rightarrow0]{}{\longrightarrow}1,\\
&\partial_t\hat{p}^{(0)}(k)\stackrel[k\rightarrow0]{}{\longrightarrow}0,\\
&\partial_{tt}\hat{p}^{(0)}(k)\stackrel[k\rightarrow0]{}{\longrightarrow}0,
\end{align}
\end{subequations}
where the first limit is satisfied by normalization and the last two can be computed directly from the equations \eqref{IniCurrent} and \eqref{2partialTimeP0} in the appendix. Thus $\widetilde{p}(k;\epsilon)$ simplifies to 
\begin{equation}\label{MarginalP-FL-AD}
\widetilde{p}_\textrm{A-D}(k;\epsilon)=\frac{1}{\epsilon+\dfrac{\Theta^2}{\lambda}\bigl[\text{i} kv_\text{drift}+ k^2 D_\text{act}\bigr]},
\end{equation}
which can be written, after Fourier-Laplace inversion, as the advection-diffusion equation
\begin{equation}
 \frac{\partial}{\partial t}p_\textrm{A-D}(x;t)+
v_\textrm{eff}
\frac{\partial}{\partial x}p_\textrm{A-D}(x,t)
=D_\textrm{eff}\frac{\partial^2}{\partial x^2}p_\textrm{A-D}(x;t),
\end{equation}
with the effective drift velocity $v_\textrm{eff}$ given by
\begin{subequations}
\begin{align}\label{veff}
v_\textrm{eff}\equiv&\frac{\Theta^2}{\lambda}v_\textrm{drift}\nonumber\\
=&v(p_+^\textrm{st}-p_-^\textrm{st}),
\end{align}
and the effective diffusion coefficient $D_\textrm{eff}$ by
\begin{align}
D_\textrm{eff}\equiv&\frac{\Theta^2 D_\textrm{act}}{\lambda}\nonumber\\
=&v^2\frac{\vartheta_0^-+\vartheta_0^+}{\lambda}.\label{Deff}
\end{align}
\end{subequations}
Thus in the long-time, large-distance regime the probability density of the particle position is a traveling Gaussian given by
\begin{equation}\label{eq:Gaussian}
p_\textrm{A-D}(x;t)\sim\frac{1}{\sqrt{4\pi D_\textrm{eff}t}}\exp\biggl\{-\frac{(x-v_\textrm{eff}t)^2}{4 D_\textrm{eff}t}\biggr\}.
\end{equation}
We thus expect the kurtosis, $\kappa$, of such distribution to acquire the value 3.
The first moments in this approximation are known to be 
\begin{subequations}
\begin{equation}
\langle x(t)\rangle_{p_{A-D}}=v_\textrm{eff}t
\end{equation}
 and 
\begin{equation}
\langle x^2(t)\rangle_{p_{A-D}}=v^2_\textrm{eff}t^2+2D_\textrm{eff}t.
\end{equation}
\end{subequations}
Notice that two terms contribute to the mean squared displacement in this long-time regime, a ballistic one originated in the non-reciprocity of the transition between self-propulsion states and a diffusive one. The case $v_\textrm{drift}=0$ makes $v_\textrm{eff}=0$, and thus only diffusive transport is present. As was anticipated in the previous section, the equation $v_\textrm{drift}=0$ defines a manifold in the space of the transition rates which for which normal diffusion with diffusion coefficient $D_\textrm{eff}$ is is observed. Different cases will be thoroughly analyzed in the following section. 

%------------
\paragraph{Short-time and small-distance regimes}
In the case of short times ($\epsilon\gg\Theta$) and small lengths, ($k\gg\Theta/v$), $\widetilde{p}(k;\epsilon)$ given in \eqref{MarginalP-FL} can be approximated by
\begin{multline}\label{Sol-ST}
\widetilde{p}(k;\epsilon)\approx\frac{1}{\epsilon}\Bigl[\hat{p}^{(0)}(k)+\frac{\epsilon\, \partial_t\hat{p}^{(0)}(k)}{\epsilon^2+k^2v^2}\\
+\frac{\Theta\partial_t\hat{p}^{(0)}(k)+\partial_{tt}\hat{p}^{(0)}(k)}{\epsilon^2+k^2v^2}\Bigr],
\end{multline}
which differs from the standard wave-like behavior of the well-known two-state run-and-tumble model (obtained  from Eq. \eqref{3stat-masteq} by setting $\vartheta_0^+=\vartheta_0^-=\vartheta_+^0=\vartheta_-^0=0$). In such a  case we recall that the solution in the short-time regime is approximated by 
\begin{equation*}
\widetilde{p}_\textrm{w}(k;\epsilon)=\frac{\epsilon\,\hat{p}^{(0)}_\textrm{w}(k)+\partial_{t}\hat{p}^{(0)}_\textrm{w}(k)}{\epsilon^2+k^2v^2},
\end{equation*}
which after inversion of the Fourier-Laplace transform, leads to the well-known  D'Alambert solution of the one-dimensional wave equation
\begin{multline}\label{D'Alambert}
p_\textrm{w}(x;t)=\frac{1}{2}\Bigl[p^{(0)}_\textrm{w}(x+v\,t)+p^{(0)}_\textrm{w}(x-v\,t)\Bigr]\\
+\frac{1}{2v}\int\limits_{x-vt}^{x+vt}dx^\prime\, \partial_tp^{(0)}_\textrm{w}(x^\prime),
\end{multline}
whose first term encodes the traveling of the initial distribution, half to the left with speed $-v$ and half to the right with speed $v$. The second term gives the contribution of spatial change of initial probability current propagated to all positions in the interval $[x-vt,x+vt]$ up to time $t$. 

In contrast, the approximated solution given by expression \eqref{Sol-ST} leads, after taking the direct inverse Fourier transform, to
\begin{multline}
\label{ShortTime-Approx}
p(x,t)\approx p^{(0)}(x)+\frac{1}{2v}\int_{x-vt}^{x+vt}\hspace{-0.2cm}dx^\prime\, \partial_t p^{(0)}(x^\prime)\\
+\frac{\Theta t}{2v}\int_{x-vt}^{x+vt}\hspace{-0.2cm}dx^\prime\, \textrm{tri}\Bigl(\frac{x-x^\prime}{vt}\Bigr)\partial_t p^{(0)}(x^\prime)\\
+\frac{t}{2v}\int_{x-vt}^{x+vt}\hspace{-0.2cm}dx^\prime\, \textrm{tri}\Bigl(\frac{x-x^\prime}{vt}\Bigr)\partial_{tt}p^{(0)}(x^\prime),
\end{multline}
where $\textrm{tri}(x)$ refers to the \emph{triangle function} defined to be $1-|x|$ if $|x|\le1$ and vanishes otherwise. In the expression \eqref{ShortTime-Approx} we notice that the second term in the right-hand side gives the same contribution as in the D'Alambert solution \eqref{D'Alambert} due to $\partial_t p^{(0)}(x)$, but that there is no contribution of the split initial pulse being propagated with velocities $\pm v$, instead, in the short-time regime, propagation is driven by the fourth term of Eq.~\eqref{ShortTime-Approx}. This is more clear if we use Eq.~\eqref{2partialTimeP0} for $\partial_{tt}p^{(0)}$(k) given in the appendix, since this leads to
\begin{widetext}
\begin{multline}
\label{ShortTime-Approx2}
p(x,t)\approx p^{(0)}_0(x)+\frac{1}{2}\Bigl[p^{(0)}_-(x-vt)+p^{(0)}_+(x-vt)\Bigr]+\frac{1}{2}\Bigl[p^{(0)}_-(x+vt)+p^{(0)}_+(x+vt)\Bigr]
+\frac{1}{2v}\int_{x-vt}^{x+vt}\hspace{-0.2cm}dx^\prime\, \partial_t p^{(0)}(x^\prime)\\
+\frac{\Theta t}{2v}\int_{x-vt}^{x+vt}\hspace{-0.2cm}dx^\prime\, \textrm{tri}\Bigl(\frac{x-x^\prime}{vt}\Bigr)\partial_t p^{(0)}(x^\prime)
+\frac{t}{2}\int_{x-vt}^{x+vt}\hspace{-0.2cm}dx^\prime\, \textrm{tri}\Bigl(\frac{x-x^\prime}{vt}\Bigr)\times\\
\frac{\partial}{\partial x^\prime}\biggl[(\vartheta_+^0+2\vartheta_+^-)p^{(0)}_+(x^\prime)
+(\vartheta_0^--\vartheta_0^+)p_0^{(0)}(x^\prime)
-(\vartheta_-^0+2\vartheta_-^+)p^{(0)}_-(x^\prime)\biggr],
\end{multline}

\begin{figure*}
\includegraphics[width=0.34\textwidth,trim=5 50 0 40,clip=true]{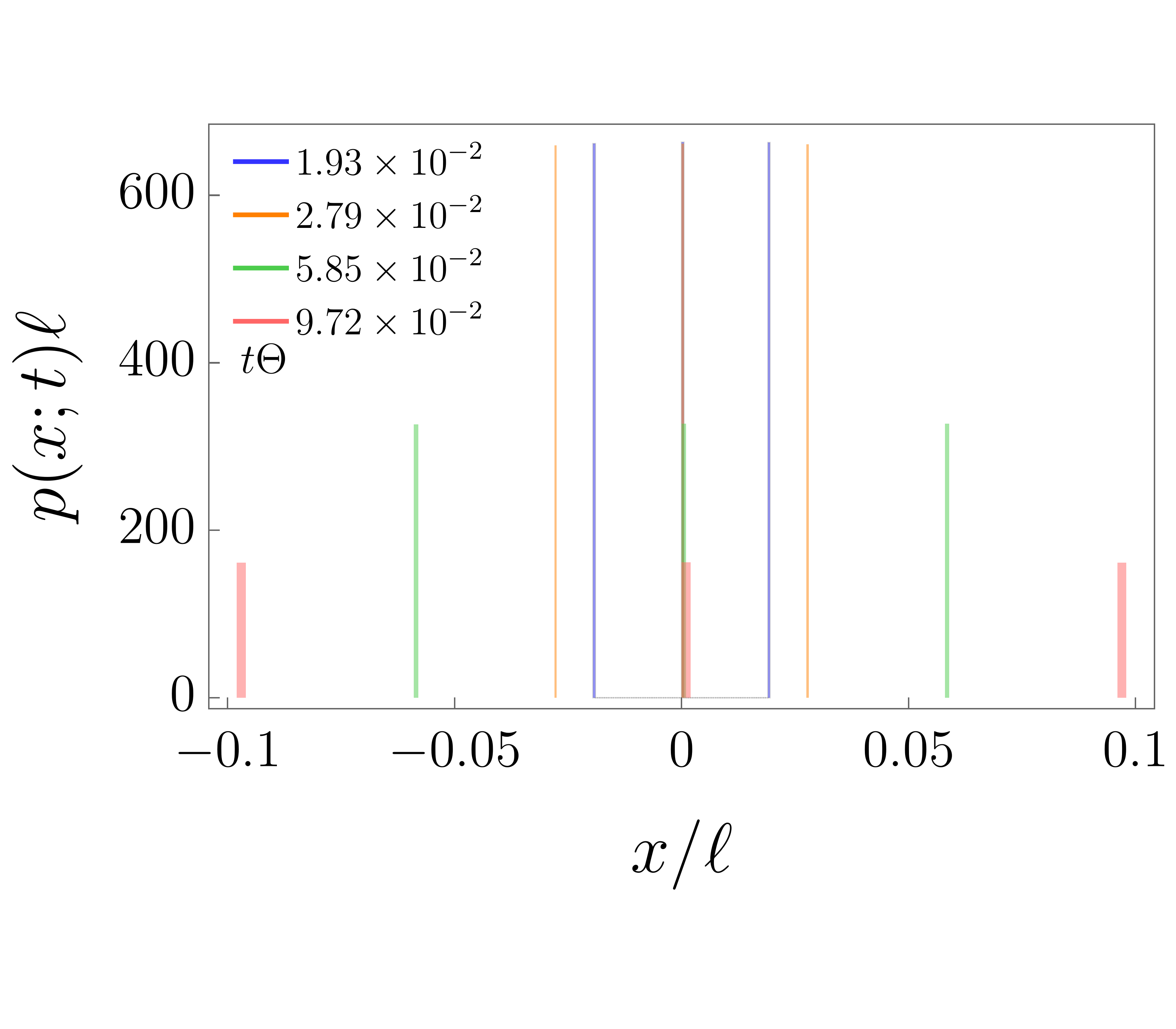}\includegraphics[width=0.3\textwidth,trim=0 30 0 0,clip=true]{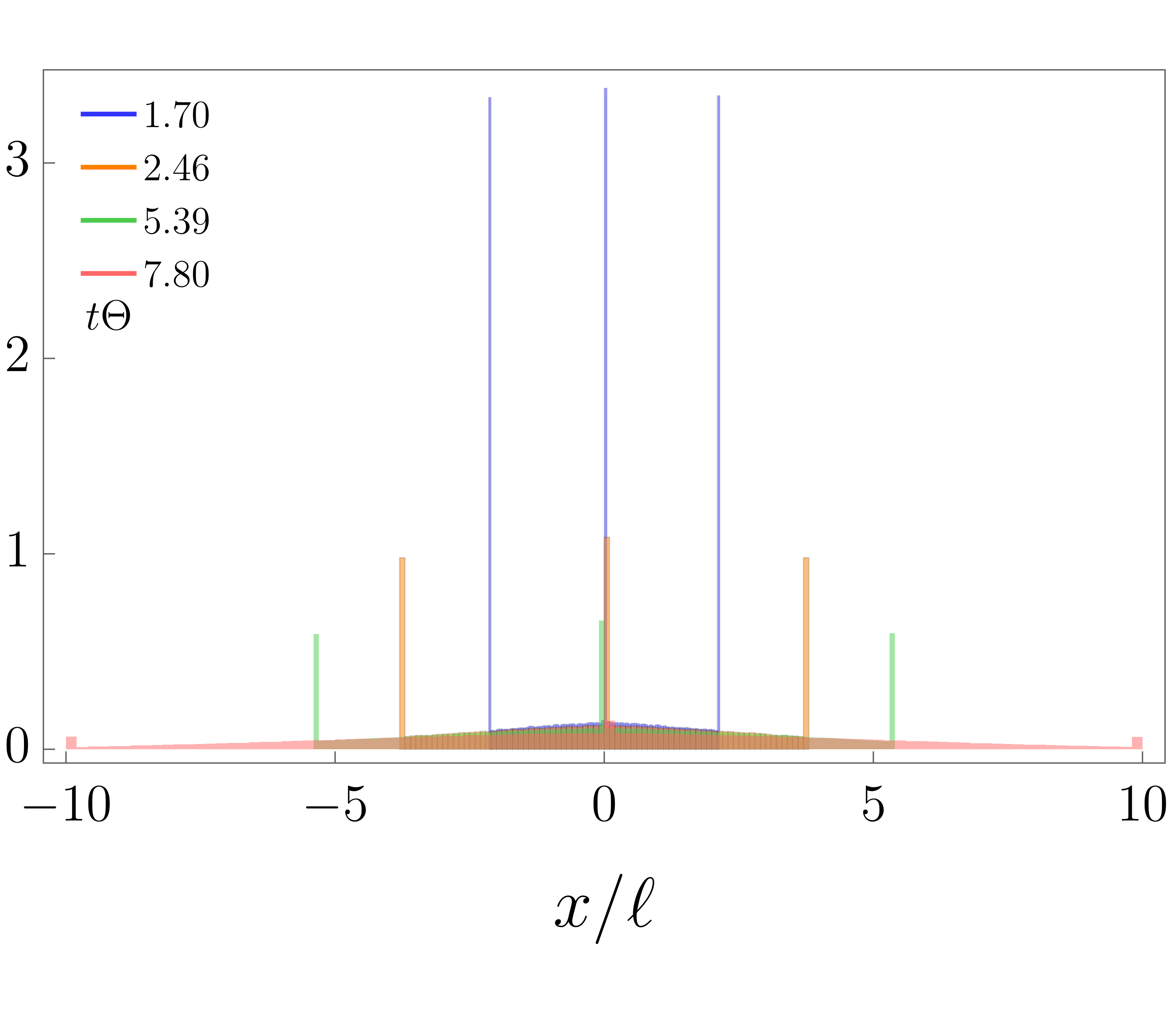} \includegraphics[width=0.325\textwidth,trim=0 45 0 10,clip=true]{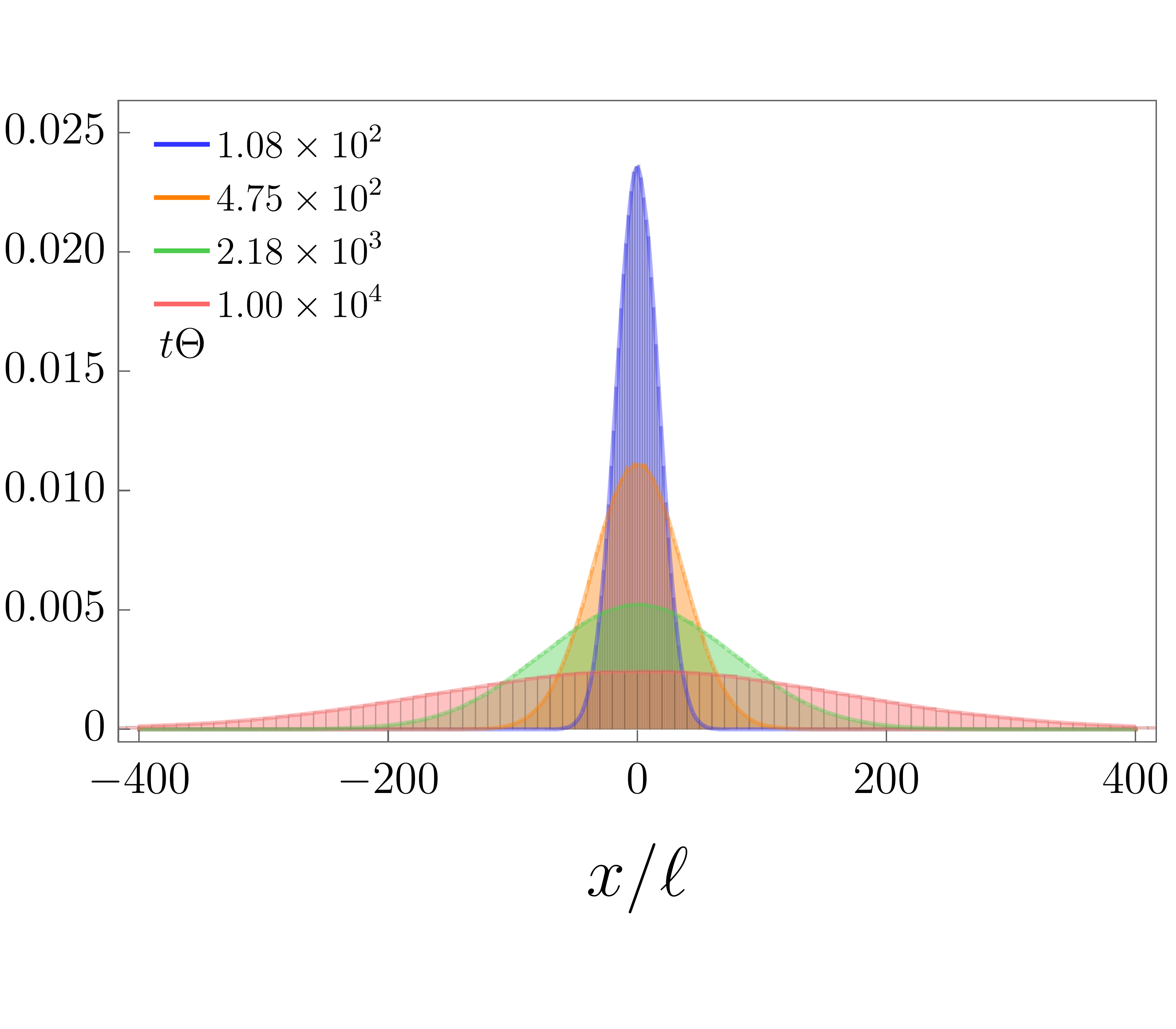}
\caption{The dimensionless probability density $p(x;t)\ell$ obtained numerically from  an ensemble of $3\times10^5$ trajectories for the initial conditions $p_-^{(0)}(x)=p_0^{(0)}(x)=p_+^{(0)}(x)=\frac{1}{3}\delta(x)$ and reciprocal and homogeneous transitions rates, $\vartheta_n^m=\Theta/6$. Left panel.- Short-time regime, $t\Theta\ll1$, where the three contributions to $p(x;t)$ in Eq.~\eqref{pdf-short}  are clearly shown. Central panel.- Intermediate-time regime, $t\Theta\simeq1$, the spreading of the pulses is clearly shown. Right panel.- The long-time regime, $t\Theta\gg1$ sets in, the positions distribution is well described by a Gaussian distribution.}
\label{fig:histogram_symmetric_rates}
\end{figure*}
\end{widetext}
where the first line is reminiscent of the D'Alambert solution of the one-dimensional wave-equation, the next terms give the contributions to the propagation that are not described by the wave equation.

In the case for which the initial distributions of the self-propulsion states are equally distributed and spatially concentrated as a pulse in the origin, i.e. $p_-^{(0)}(x)=p_0^{(0)}(x)=p_+^{(0)}(x)=\frac{1}{3}\delta(x)$ we have $\partial_t p^{(0)}(x)=0$ since we have vanishing initial probability current, and if in addition the transitions rates are all equal $\vartheta_n^m=\Theta/6$, we have that  $\partial_{tt}p^{(0)}(x)=0$. In such a case we get 
\begin{multline}
\label{pdf-short}
p(x,t)\approx p^{(0)}_0(x)+\frac{1}{2}\Bigl[p^{(0)}_-(x-vt)+p^{(0)}_+(x-vt)\Bigr]\\
+\frac{1}{2}\Bigl[p^{(0)}_-(x+vt)+p^{(0)}_+(x+vt)\Bigr],
\end{multline}
where the distribution of the particles in the ensemble that initiates at state of rest ($v=0$) can be clearly identified along with the two symmetric pulses traveling to the right and left with speed $\pm v$, this contribution is reminiscent of the wave propagation. Last expression is in correspondence with the results obtained from numerical simulations---see left panel of Fig.~\ref{fig:histogram_symmetric_rates}---where each contribution to Eq.~\eqref{pdf-short} is clearly shown at different times $t\Theta^{-1}\ll1$. 

%%%
\subsection{The Intermediate Scattering Function}
The probability density, $\hat{p}(k;t)$, the Fourier transform of the probability distribution of the particle displacements is known as the \emph{intermediate scattering function} (ISF), this has been analyzed theoretically \cite{KurzthalerSciRep2016,KurzthalerSoftMatter2017,Breoni2022} and experimentally, in active matter systems \cite{KurzthalerPRL2018}. This encodes full statistical information about the particle displacements at a given length scale $k^{-1}$ and delay time $t$. This is in general a real valued quantity that relaxes monotonically to zero with time when the particle displacements are short-term correlated, contrarily, slight oscillations are observed in the case of correlated displacements.   

In our case the ISF is obtained from \eqref{MarginalP-FL} after inverting the Laplace transform. Since the analytical inversion is a rather difficult task, we carry it out numerically.  
\begin{figure}
\includegraphics[width=0.9\columnwidth]{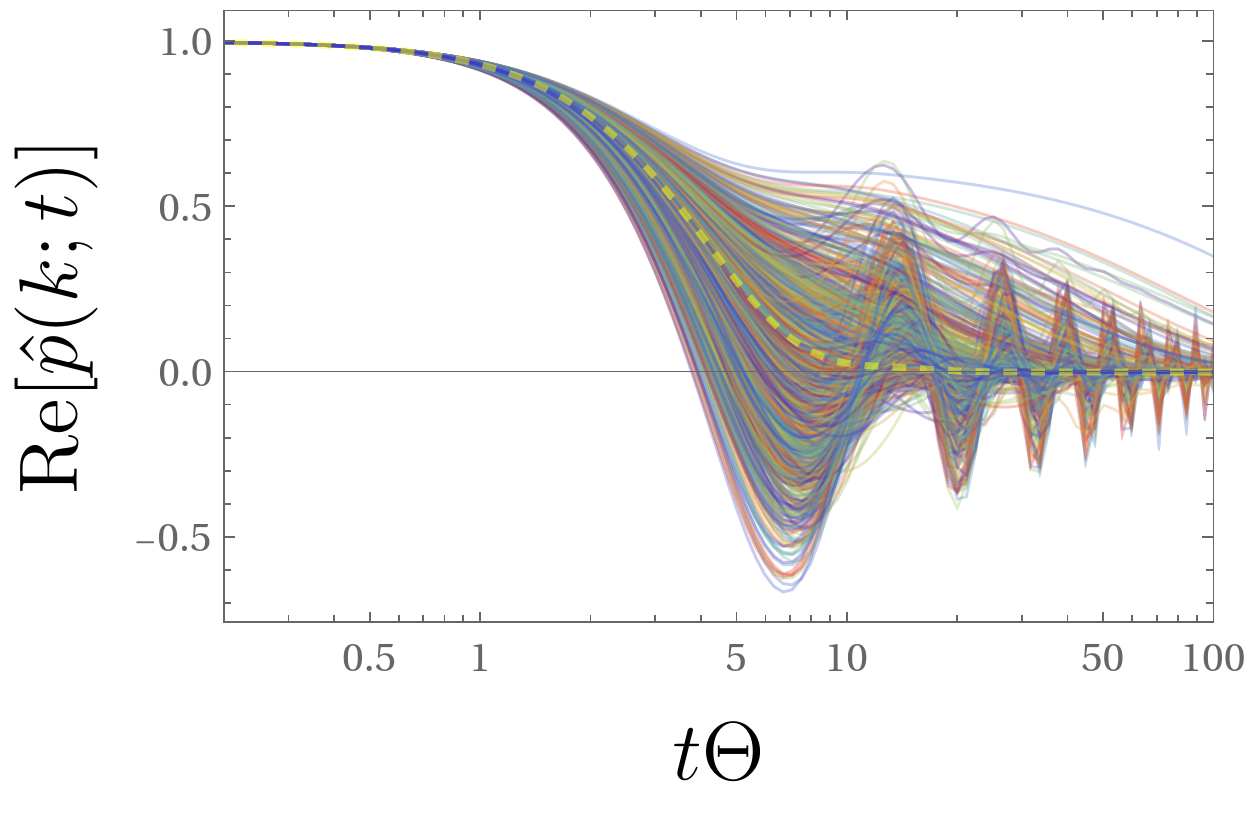}
\includegraphics[width=0.9\columnwidth]{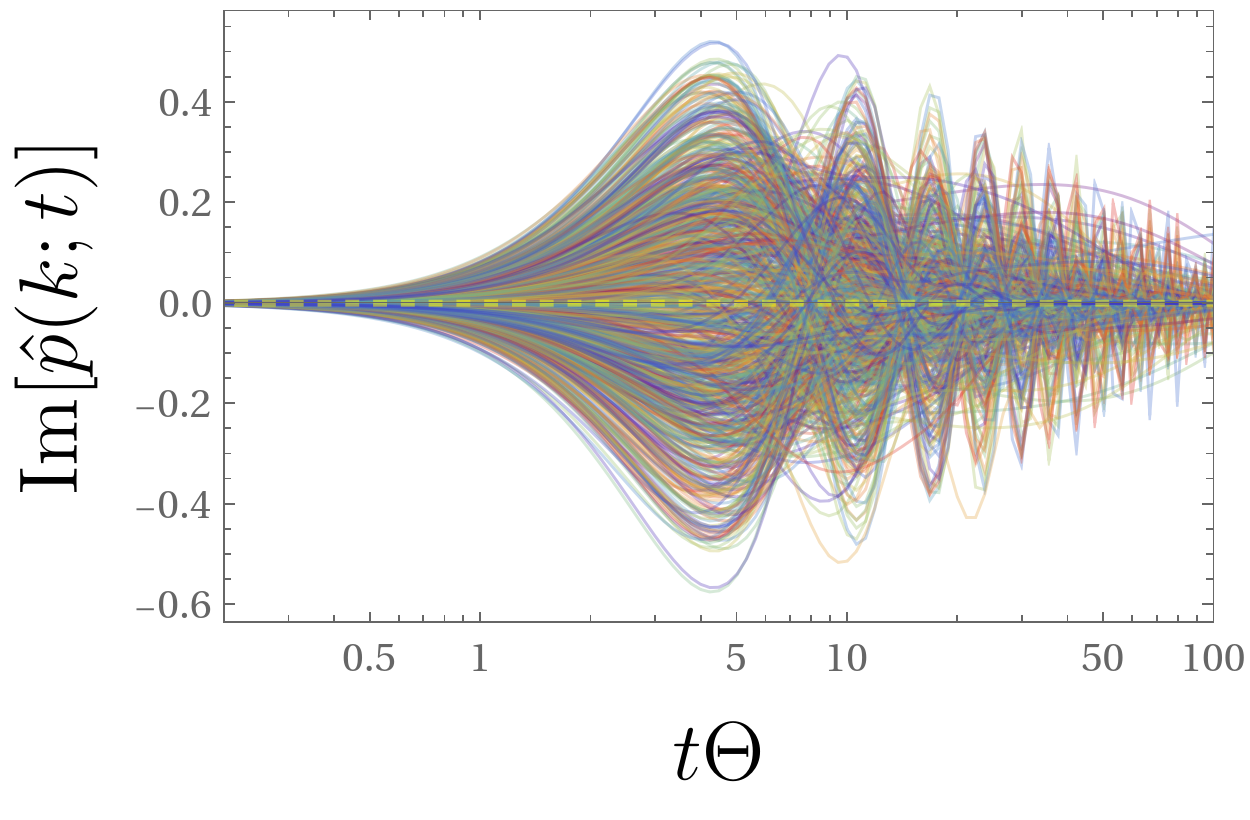}
\caption{The time dependence of the intermediate scattering function $\hat{p}(k;t)$, for $k=0.5\ell^{-1}$ and $10^3$ six-tuples uniformly sampled from the 5-simplex $\mathbb{H}$. Top panel.- The real part 0f $\hat{p}(k;t)$ is shown. Bottom panel.- The imaginary part of $\hat{p}(k;t)$ is shown. Dashed-yellow lines mark the reciprocal- transitions case $\widetilde{\vartheta}_n^m=\frac{1}{6}$.}
\label{fig:ISF}
\end{figure}
In Fig.~\ref{fig:ISF} the time dependence of the real and imaginary parts of the ISF are shown as function of time for $k=0.5\ell^{-1}$ (which encodes the statistics of the particle displacements of the order of the characteristic length $\ell$), for which $10^3$ six-tuples $\{\widetilde{\vartheta}_-^+,\widetilde{\vartheta}_-^0,\widetilde{\vartheta}_0^-,\widetilde{\vartheta}_0^+,\widetilde{\vartheta}_+^0, \widetilde{\vartheta}_+^-\}$ uniformly samples in the 5-simplex were considered. The intersection of the manifolds $v_\textrm{drift}=0$ and $v_\textrm{act}=0$ in the space of transition rates space make the $\hat{p}(k;t)$ a real valued function, as is shown in particular for the completely reciprocal case (dashed-yellow lines). Strongly coherent motion (ballistic transport) is induced by transitions rates out of this intersection, as is revealed in the strong oscillations in real part of $\hat{p}(k;t)$ and in its non-vanishing imaginary part. In the case of the results presented in Ref. \cite{Breoni2022}, such strong coherent motion is induced by the unequal self-propelling velocities chosen.

%%%%%%
\section{Moments of $p(x;t)$}
\label{Sect:Moments}
The explicit solution \eqref{MarginalP-FL} in Fourier-Laplace variables,  provides the Laplace transform of the $n$-th moment of the position distribution $p(x;t)$, $\langle x^n(t)\rangle_\textrm{act}$, that carries out the statistical information of the active process through the formula
\begin{align}\label{Def:Moments}
\bigl\langle\widetilde{x^n}(\epsilon)\bigr\rangle_\textrm{act}&=\int_{-\infty}^\infty dx\, x^n \mathcal{L}\{p(x;t)\}\nonumber\\
&=\textrm{i}^{n}\frac{\partial^{n}}{\partial k^{n}}\widetilde{p}(k;\epsilon)\biggr|_{k=0}. \end{align}
From this, $\bigl\langle x^n(t)\bigr\rangle_\textrm{act}$ is obtained after taking the inverse Laplace transform. In the following we focus our discussion on the first moments. We compare our analytical results with numerical ones obtained for $10^3$ six-tuples of transitions rates chosen uniformly at random from the 5-simplex $\mathbb{H}$. For each six-tuple, the first moments were computed from an ensemble of $3\times10^5$ trajectories generated from the integration of Eq.~\eqref{Langevin} using Kinetic Monte Carlo methods. All simulations were carried out with the initial distributions 
\begin{equation}
\label{eq:IniDistributions}
p_-(x;0),\; p_0(x;0),\; p_+(x;0)=\frac{1}{3}\delta(x).
\end{equation}

\subsection{The mean position $\langle x(t)\rangle_\textrm{act}$}

The contribution of active motion to the mean position of the particle in the Laplace variable can be computed directly from \eqref{Def:Moments} with $n=1$
\begin{multline}\label{FirstMomentLaplace}
\bigl\langle \widetilde{x}(\epsilon)\bigr\rangle_\text{act} = \frac{\bigl\langle x(0)\bigr\rangle_\textrm{act}}{\epsilon}+v_\textrm{drift} \frac{\widetilde{F}(0;\epsilon)}{\epsilon^2}\\
+\left(\frac{J^{(0)}}{\Theta}+\frac{\alpha}{\Theta^2}\right)\frac{\widetilde{F}(0;\epsilon)}{\epsilon}
+\frac{J^{(0)}}{\Theta^2}\widetilde{F}(0;\epsilon),
\end{multline}
where we have used that $\hat{p}^{(0)}(k)\big\vert_{k=0}=1$ and $\bigl[\partial_t\hat{p}^{(0)}(k)\bigr]_{k=0}=0$, $\bigl[\partial^2_t\hat{p}^{(0)}(k)\bigr]_{k=0}=0$ as can be checked from \eqref{IniCurrent} and \eqref{DotIniCurrent} in the appendix; $J^{(0)}$ is the total initial probability current as defined in \eqref{TotalCurrent}; and $\alpha$ is computed from initial conditions given in \eqref{eq:alpha}. The Laplace inversion of \eqref{FirstMomentLaplace} can be written as
\begin{multline}
\bigl\langle x(t)\bigr\rangle_\text{act} =\bigl\langle x(0)\bigr\rangle_\textrm{act}+ v_\textrm{drift}\int_0^t ds\int_0^s ds_1\,  F(0;s_1)\\
+\left(\frac{J^{(0)}}{\Theta}+\frac{\alpha}{\Theta^2}\right)\int_0^t ds\, F(0;s)\\
+J^{(0)}F(0;t),
\end{multline}
where 
\begin{equation}
F(0;t)=\Theta^2\, e^{-\Theta t/2}
\begin{cases}
\dfrac{\sin(\vert\Delta\vert t)}{\vert\Delta\vert}&\textrm{if}\; \Delta^2>0\\
\qquad t & \textrm{if}\; \Delta=0\\ 
\dfrac{\sinh(\vert\Delta\vert t)}{\vert\Delta\vert}&\textrm{if}\; \Delta^2<0
\end{cases},
\end{equation}
is the inverse Laplace transform of
\begin{equation}
\widetilde{F}(0;\epsilon)=\dfrac{\Theta^2}{\bigl(\epsilon+\frac{\Theta}{2}\bigr)^2+\Delta^2}.
\end{equation}
Typical values of $\Delta^2$ are non positive as is shown in Fig.~\ref{fig:Delta2}, for these we have $F(0;t)\sim e^{-\Theta t(1/2-\vert\Delta\vert)}$ for $\Theta t\gg1$ ($\vert\Delta\vert<\frac{1}{2}$), this decays with time at a slower pace than $e^{-\Theta/t2}$ for $\vert\Delta\vert>0$.

\begin{figure}
\includegraphics[width=1.0\linewidth]{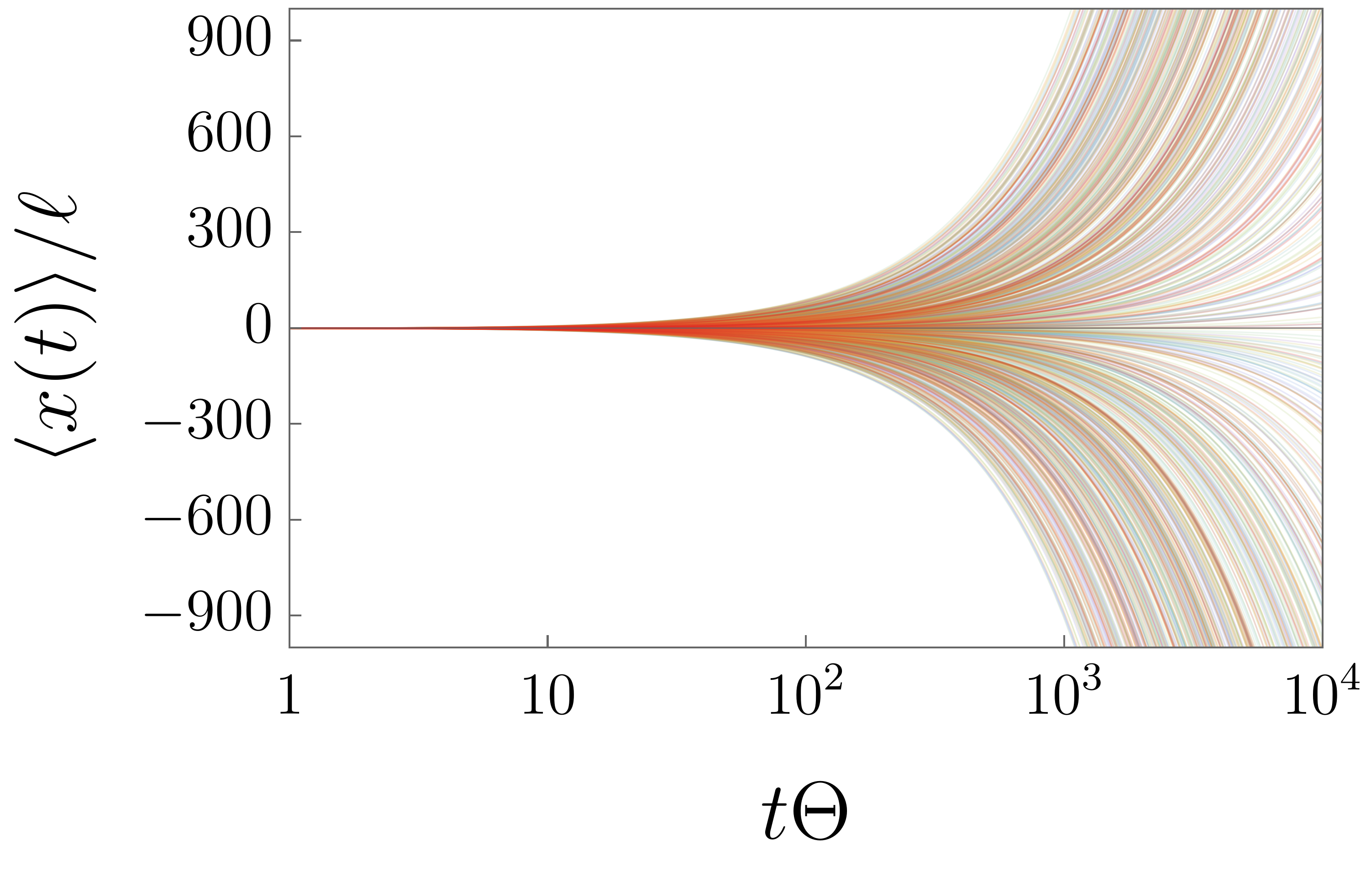}
    \caption{The dependence of the mean position, $\langle x(t)\rangle_\textrm{act}/\ell$, on the dimensionless time $t\Theta$, is shown for a sample of $10^3$ six-tuples of transitions rates. Each curve was computed numerically from an ensemble of $3\times10^5$ trajectories generated from the solutions of \eqref{Langevin}.
    }
    \label{fig:gral_mean}
\end{figure}

In the homogeneous and reciprocal case, the mean positions vanishes under initial conditions that correspond to a pulse at the origin with vanishing probability current.   
In the typical case we have that the time dependence of the mean position transits from  an ``accelerated'' regime at the short-time limit \begin{equation}\label{m1:short-times}
  \bigl\langle x(t)\bigr\rangle_\text{act} \approx \bigl\langle x(0)\bigr\rangle_\text{act} + J^{(0)}\,t+ \dfrac{\alpha\,t^2}{2},
\end{equation}
to a ballistic regime in the long-time-regime 
\begin{equation}\label{m1:long_times}
    \bigl\langle x(t)\bigr\rangle_\text{act} \sim v_\textrm{eff}\,t.
\end{equation}
This process is reminiscent of the motion of a particle under a constant force field (per unit mass) $\alpha$, which is effectively  dissipated towards a ballistic motion with a terminal speed $v_\textrm{eff}$ given in \eqref{veff}. The process originates solely from the non-reciprocity in the transition rates between internal states, as $\alpha=0$ only if all transition rates are mutually reciprocal and the initial probability distribution of self-propulsion states is uniform. In Fig.~\ref{fig:gral_mean} the time dependence of the mean position $\langle x(t)\rangle_\textrm{act}$ obtained form numerical simulations is shown.

%%%%%%%%%%%%%%%%%%%%%%%%%%%%
\subsection{The mean squared displacement $\langle x^2(t)\rangle_\textrm{act}$}
\label{sect:MSD}
The second moment of the distribution $p(x;t)$, which correspond to the MSD, provides information on the transport properties induced by the self-propulsion pattern of motion. From Eq.~\eqref{Def:Moments} with $n=2$ we get the Laplace transform of the MSD given explicitly by
\begin{widetext}
\begin{multline}
\label{MSD-Laplace}
\Bigl\langle \widetilde{x^2}(\epsilon)\Bigr\rangle_\text{act} = \frac{1}{\epsilon}\bigl\langle x^2(0) \bigr\rangle_\textrm{act}  + 
\frac{2}{\epsilon^2}
\biggl[v_\textrm{eff}\,\bigl\langle x(0)\bigr\rangle_\textrm{act}+D_\textrm{eff}+\dfrac{v_\textrm{eff}\,v_\textrm{act}}{\Theta}\widetilde{F}(0;\epsilon)\biggr]
+
\dfrac{2\, v_\textrm{eff}^2}{\epsilon^3} +
\\ 
+ \widetilde{F}(0;\epsilon)\frac{1}{\Theta}\left(\dfrac{\gamma}{\epsilon} +\dfrac{\gamma}{\Theta} + \dfrac{\beta}{\epsilon\, \Theta}\right) + \frac{2\,v_\textrm{eff}}{\epsilon\Theta}\widetilde{F}(0;\epsilon) \left(\dfrac{J^{(0)}}{\epsilon} +\dfrac{J^{(0)}}{\Theta}+\dfrac{\alpha}{\Theta\,\epsilon} \right) \\
+ \frac{2v_\textrm{act}}{\Theta^2}\widetilde{F}^2(0;\epsilon) \left(\dfrac{J^{(0)}}{\Theta} +\dfrac{J^{(0)}}{\epsilon}+\dfrac{\alpha}{\Theta\,\epsilon}\right).
\end{multline}
Although the Laplace transform can, in principle, be inverted term by term, the resulting expression is too cumbersome and lengthy to clearly reveal the time dependence. Despite of this, the initial distribution \eqref{eq:IniDistributions}, allow further simplifications of expression \eqref{MSD-Laplace} ($\gamma=0$, $J^{(0)}=0$ see the appendix \ref{app:InitialConditions}), in that case we have
\begin{equation}
\label{MSD-SimpleICs}
\Bigl\langle \widetilde{x^2}(\epsilon)\Bigr\rangle_\text{act} = 
v_\textrm{eff}^2\frac{2}{\epsilon^3}+
2D_\textrm{eff}\frac{1}{\epsilon^2}
+\frac{\widetilde{F}(0;\epsilon)}{\epsilon\Theta}
\biggl[\frac{2 v_\textrm{eff}}{\epsilon}\Bigl(v_\textrm{act}+\frac{\alpha}{\Theta}\Bigr)+\frac{\beta}{\Theta}\biggr]
+\frac{\widetilde{F}^2(0;\epsilon)}{\epsilon\Theta}\frac{2v_\textrm{act}\alpha}{\Theta^2}.
\end{equation}
The inverse Laplace transform of the first two terms are straightforwardly computed and give $v_\textrm{eff}^2 t^2$ and $2D_\textrm{eff}t$, respectively.
\end{widetext}

\begin{figure}
\includegraphics[width=1.0\linewidth]{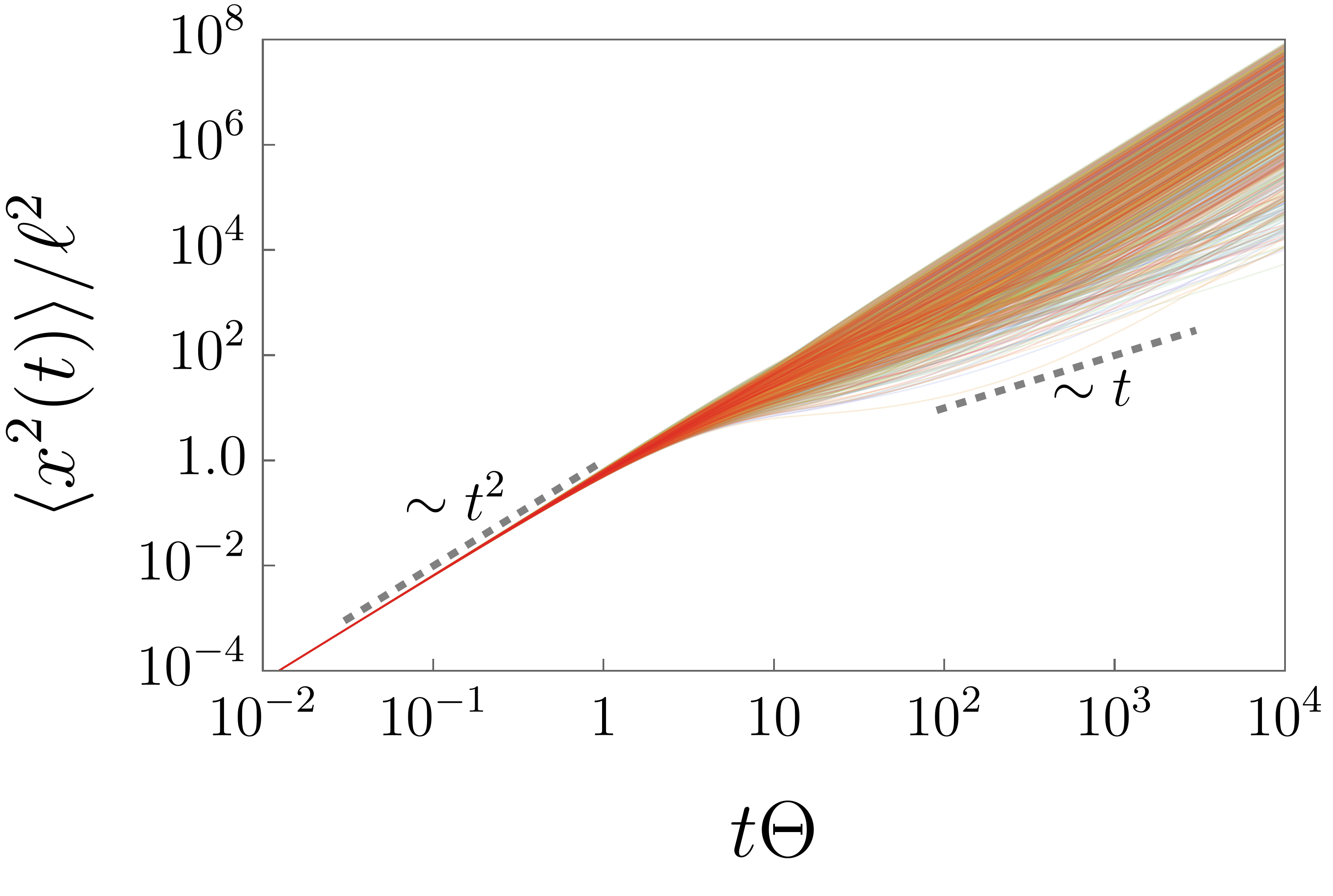}
\caption{The time dependence of the dimensionless MSD is shown as function of the dimensionless time 
 $t\Theta$ for a sample of $10^3$ six-tuples of transitions rates. Each curve was computed numerically from an ensemble of $3\times10^5$ trajectories generated from the solutions of \eqref{Langevin}. The ballistic transport regime ($v^2t^2$) is clear at short times, while at long times two regimes are noticed depending on whether $v_\textrm{eff}$ vanishes or not, a diffusive regime in the first case ($2D_\textrm{eff}\, t$) and ballistic $v_\textrm{eff}^2t^2$ in the other. The spreading of the curves in the long-time regime indicates the influence of the six-tuple chosen values on $v_\textrm{eff}$ \eqref{veff} and $D_\textrm{eff}$ \eqref{Deff}.
 }
    \label{fig:gral_MSD}
\end{figure}

In the short-time regime ($t\Theta\ll1$ equivalently $\epsilon/\Theta\gg1$), it can be shown the expected ballistic propagation to appear 
\begin{subequations}
\begin{equation}
\label{eq:MSD-ShortTimes}
\bigl\langle x^{2}(t)\bigr\rangle_\text{act} \approx \frac{1}{2}\beta t^2=\frac{2}{3}v^2 t^2,
\end{equation}
where we have compute $\beta$ from Eq.~\eqref{def:beta} for the initial considered. For long times ($t\Theta\gg1$, $\epsilon/\Theta\ll1$) we have
\begin{equation}
\bigl\langle x^{2}(t)\bigr\rangle_\text{act} \sim v^2_\textrm{eff}\,t^2+O(t) 
\end{equation}
where the ballistic term is the most dominant for transitions rates that make $v_\textrm{drift}\neq0$, corrected by terms that are linear in time. The standard diffusive regime is recovered on the transition-rates manifold defined by $v_\textrm{drift}=0$ for which $v_\textrm{eff}=0$ and 
\begin{equation}
\bigl\langle x^{2}(t)\bigr\rangle_\text{act} \sim 2D_\textrm{eff}t.
\end{equation}
\end{subequations}
as has been anticipated in the asymptotic analysis presented in Sect.~\ref{subsect:LongShortRegimes}. In Fig.~\ref{fig:gral_MSD} we present the time dependence of the MSD for $10^{3}$ six-tuples uniformly chosen in 5-simplex $\mathbb{H}$ and for the initial conditions precised in Eq.~\eqref{eq:IniDistributions}. In the figure, the short-time ballistic transport regime \eqref{eq:MSD-ShortTimes} is conspicuously clear (the dashed line marking the $t^2$ dependence is \eqref{eq:MSD-ShortTimes}), while in the long-time regime the time dependence is clearly either ballistic ($\sim t^2$) or normal diffusive ($\sim t$).

%%%%%%%%%%%%%%%%%%%%%%%%%%%%%%%%%%%%%%%%
\subsection{The Kurtosis of $p(x;t)$}

The kurtosis $\kappa$ of the position distribution $p(x;t)$ is given by
\begin{equation}\label{eq:kurtosis}
\kappa(t)=\frac{\bigr\langle (x-\langle x\rangle_\textrm{act})^4\bigl\rangle_\textrm{act}}{\bigr\langle (x-\langle x\rangle_\textrm{act})^2\bigl\rangle_\textrm{act}^2}.
\end{equation}
This offers a measure for the departure from the ubiquitous Gaussian distribution for which the kurtosis has the characteristic value 3, clearly this value does not identifies uniquely the Gaussian distribution \cite{WestfallAmStat2014}. 

\begin{figure}
    \centering
    \includegraphics[width=1.0\linewidth]{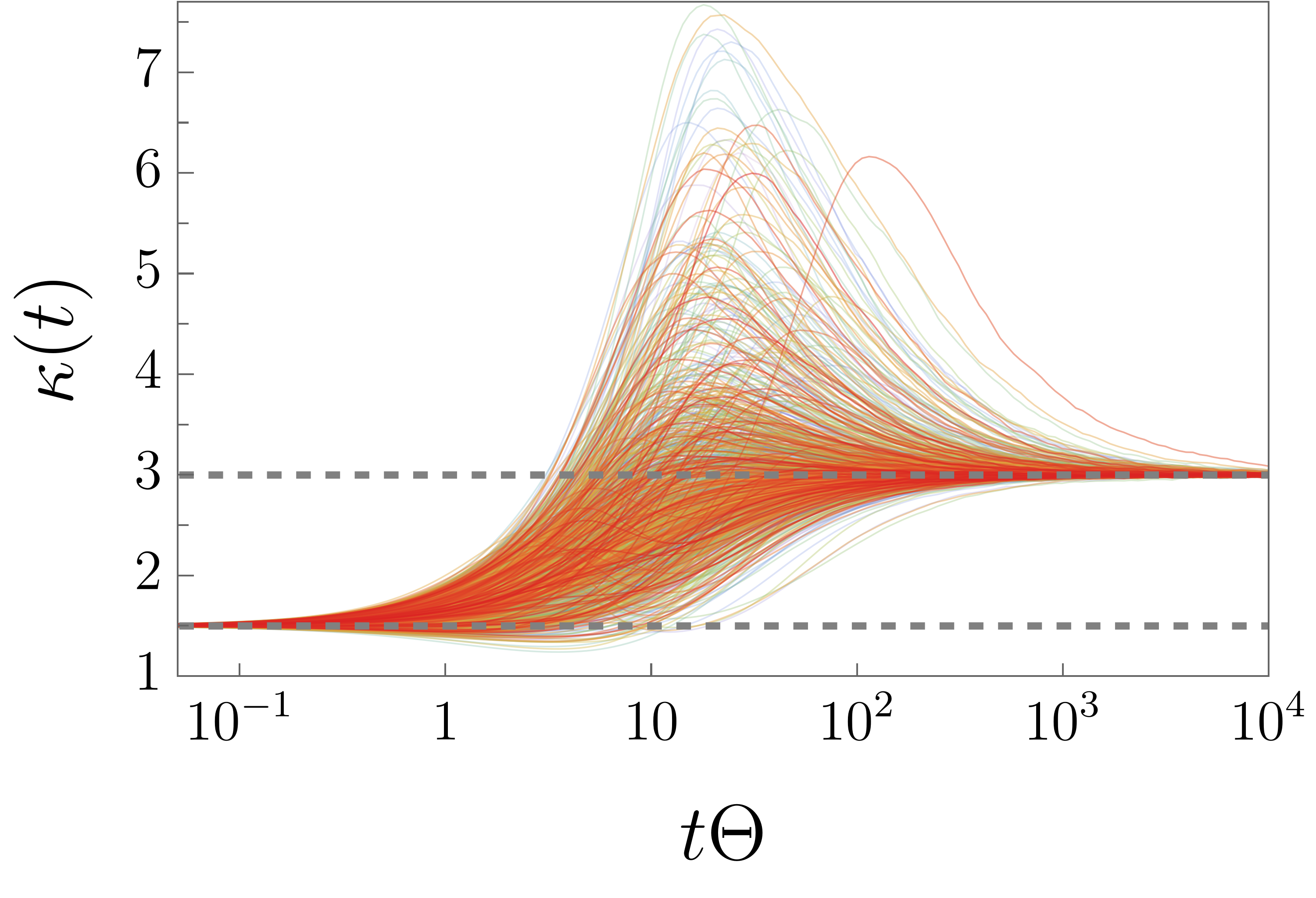}
    \caption{The time dependence of the kurtosis obtained from numerical simulations (see text) is shown.
    Horizontal dashed lines mark the values $\kappa = 1.5$ and $\kappa = 3$ that correspond to the initial distribution considered (see text), and the long-time distribution that converges to a Gaussian, respectively.
    }
    \label{fig:gral_krt}
\end{figure}

The analytical expression for the fourth moment of $p(x;t)$ is rather cumbersome and we omit it here, however, as will be shown in the next section for specific cases, a comparison between the analytical result and those results obtained from numerical simulations, are in perfect agreement. In Fig.~\ref{fig:gral_krt}  
we present $\kappa(t)$ obtained from our numerical simulations. Different time evolutions of $\kappa(t)$ are observed, however invariably it starts at $\kappa=1.5$ due to the initial conditions chosen (three equally weight pulses, see left panel of Fig.~\ref{fig:histogram_symmetric_rates}), and relaxes to $\kappa\rightarrow3$ unequivocally indicating a Gaussian distribution as was shown in Eq.\eqref{eq:Gaussian} (see right panel of Fig.~\ref{fig:histogram_symmetric_rates}). In the intermediate-time regime the specific six-tuples of the transition rates detail the time evolution of the kurtosis.

%%%%%%%%%%%%%%%%%%%%%%%%%%%%%%%%%%%%%%%%%%%%%%%%%%%%%%%%%%%%

\section{Transport properties for specific cases}
\label{Sect:ParticularCases}
\begin{figure}[h]
\includegraphics[width=0.8\columnwidth]{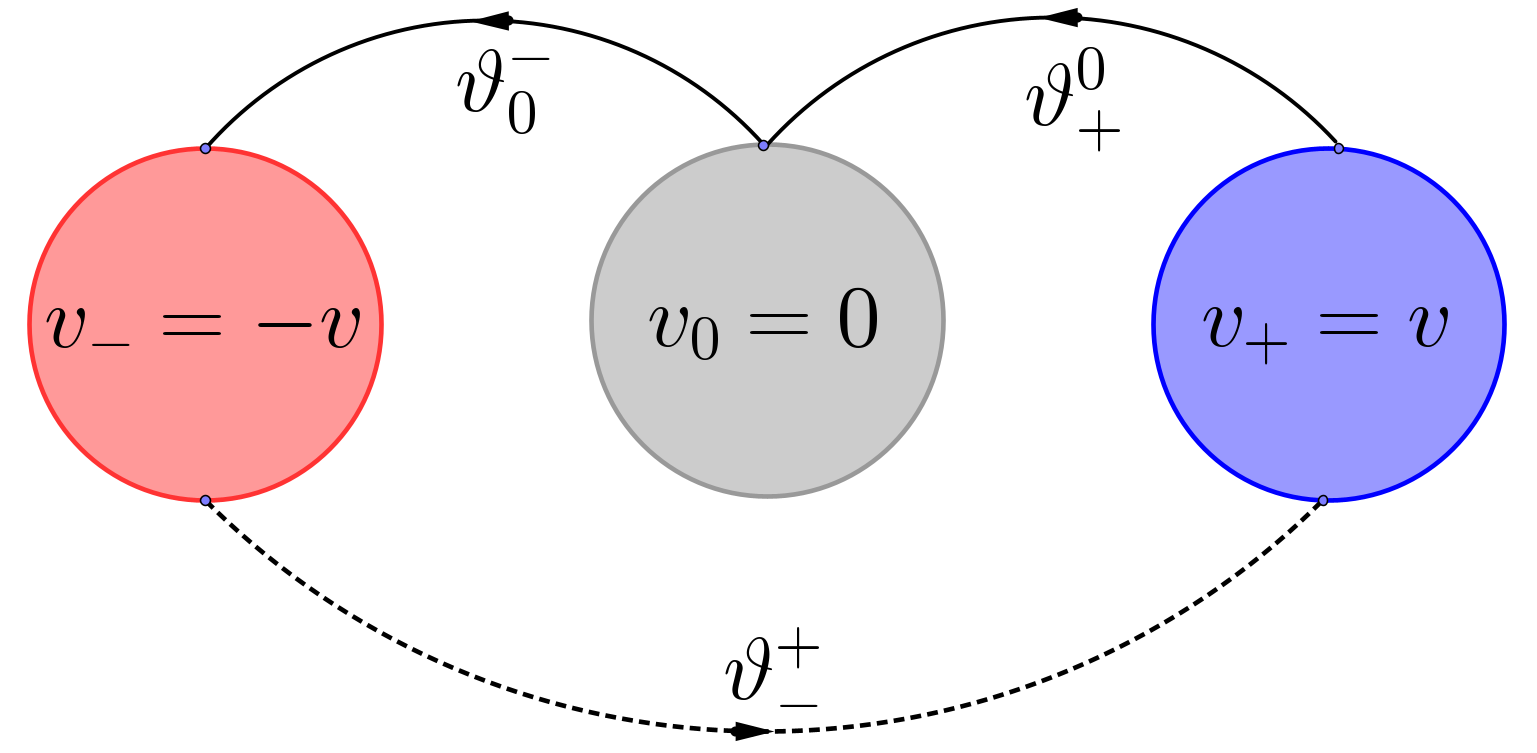}
\includegraphics[width=0.8\columnwidth,trim=0 0 0 10,clip=true]{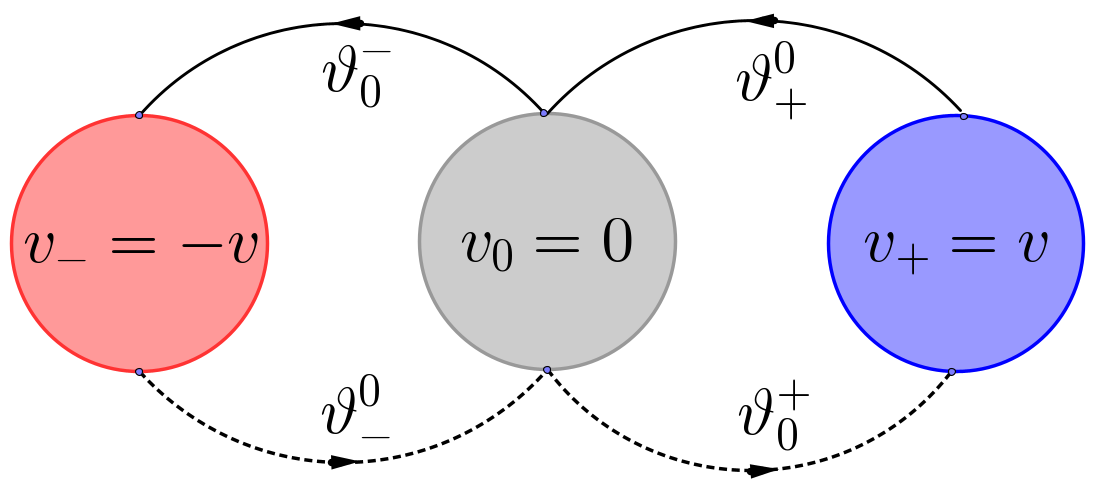}
\caption{Specific connected cyclic process considering three transitions (top panel) and four transitions (bottom panel).}
\label{fig:Transition_non-adj_rates}
\end{figure}

The non-reciprocity in our model allows for the analysis of the transport properties of different specific cases. In particular, those cases for which specific transitions do not occur. These are obtained by removing some of the connections between the self-propulsion states in Fig.~\ref{fig:Transition_rates}, i.e., by setting some of the transitions rates to the value zero. In this section, we present an explicit analysis of the cases where transitions between self-propulsion states form a connected cyclic process---i.e., configurations in which each state has at least one incoming and one outgoing transition---considering systems with three and four transition rates, respectively (see Fig.~\ref{fig:Transition_non-adj_rates}).  

%%%%%%%%
\subsection{Three transitions rates process}\label{sect-3cycle}

The simplest non-trivial case for which the non-reciprocity can be analyzed in detail corresponds to the case when only three transitions are possible. From the 20 possible configurations with three non-vanishing transitions rates (triangles in the 5-simplex), 12 configurations lead to stationary distributions of the self-propulsion states, Eqs.~\eqref{Stat}, where only one self-propulsion state is populated, and the transport properties are either ballistic when any of the states $p_\mp^\textrm{st}$ is the only one populated, or non-moving if $p_0^\textrm{st}$ is the populated state. If we denote the self-propulsion stationary distribution as the triad $p^\textrm{st}\coloneqq(p_-^\textrm{st},p_0^\textrm{st},p_+^\textrm{st})$ there are 4 triangles that leads to $p^\textrm{st}=(1,0,0)$, another four that leads to $p^\textrm{st}=(0,0,1)$ and other distinct 4 configurations to $p^\textrm{st}=(0,1,0)$. There are 4 triads of transitions rates for which only one self-propulsion state is empty, namely: the two triads $\{\vartheta_-^+,\vartheta_0^+,\vartheta_+^0\}$, $\{\vartheta_-^0,\vartheta_0^+,\vartheta_+^0\}$ lead to 
\begin{equation*}
p^\textrm{st}=\Bigl(0,\frac{\vartheta_+^0}{\vartheta_+^0+\vartheta_0^+},\frac{\vartheta_0^+}{\vartheta_+^0+\vartheta_0^+}\Bigr),
\end{equation*}
while the pair of triads $\{\vartheta_-^0,\vartheta_0^-,\vartheta_+^-\}$, $\{\vartheta_-^0,\vartheta_0^-,\vartheta_+^0\}$ lead to
\begin{equation*}
p^\textrm{st}=\Bigl(0,\frac{\vartheta_+^0}{\vartheta_+^0+\vartheta_0^+},\frac{\vartheta_0^+}{\vartheta_+^0+\vartheta_0^+}\Bigr).
\end{equation*}
These cases give rise to ballistic transport in the long-term regime.

Finally, there are four triads whose long-time transport behavior can be either ballistic or diffusive: the triads $\{\vartheta_-^+,\vartheta_0^-,\vartheta_+^-\}$, $\{\vartheta_-^+,\vartheta_0^+,\vartheta_+^-\}$ that lead to 
\begin{equation*}
p^\textrm{st}=\Bigl(\frac{\vartheta_+^-}{\vartheta_-^++\vartheta_+^-},0,\frac{\vartheta_-^+}{\vartheta_-^++\vartheta_+^-}\Bigr),
\end{equation*}
which corresponds to an effective asymmetric two states run-and-tumble system, diffusive long-time behavior is observed $v_\textrm{drift}=0$ whenever $\vartheta_+^-=\vartheta_-^+$. We focus our analysis in the triads 
\begin{equation*}
\{\vartheta_-^+,\vartheta_+^0,\vartheta_0^-\},\, \{\vartheta_-^0,\vartheta_0^+,\vartheta_+^-\}
\end{equation*}
that form a closed cycle (the first one 
\begin{equation}
\label{eq:Cycle-3rates}
v_-\stackrel{\vartheta_-^+}{\longrightarrow} v_+\stackrel{\vartheta_+^0}{\longrightarrow} v_0\stackrel{\vartheta_0^-}{\longrightarrow} v_-
\end{equation}
is shown schematically in the top panel of Fig.~\ref{fig:Transition_non-adj_rates}, the second triad is obtained by reversing the cycle), these cases give the stationary probabilities distributions 
\begin{subequations}
\begin{align}
p^\textrm{st}_-&=
\frac{\vartheta_0^-\vartheta_+^0}{\vartheta_0^-\vartheta_+^0+\vartheta_-^+\vartheta_+^0+\vartheta_-^+\vartheta_0^-}\\
p^\textrm{st}_0&=
\frac{\vartheta_-^+\vartheta_+^0}{\vartheta_0^-\vartheta_+^0+\vartheta_-^+\vartheta_+^0+\vartheta_-^+\vartheta_0^-}\\
p^\textrm{st}_+&=\frac{\vartheta_-^+\vartheta_0^-}{\vartheta_0^-\vartheta_+^0+\vartheta_-^+\vartheta_+^0+\vartheta_-^+\vartheta_0^-}
\end{align}
\end{subequations}
and
\begin{subequations}
\begin{align}
p^\textrm{st}_-&=
\frac{\vartheta_0^+\vartheta_+^-}{\vartheta_0^+\vartheta_+^-+\vartheta_-^0\vartheta_+^-+\vartheta_-^0\vartheta_0^+}\\
p^\textrm{st}_0&=
\frac{\vartheta_-^0\vartheta_+^-}{\vartheta_0^+\vartheta_+^-+\vartheta_-^0\vartheta_+^-+\vartheta_-^0\vartheta_0^+}\\
p^\textrm{st}_+&=\frac{\vartheta_-^0\vartheta_0^+}{\vartheta_0^+\vartheta_+^-+\vartheta_-^0\vartheta_+^-+\vartheta_-^0\vartheta_0^+}
\end{align}
\end{subequations}
respectively. Both cases lead to diffusive long-term behavior whenever $p_-^\textrm{st}=p_+^\textrm{st}$.
\begin{figure}[h]
\includegraphics[width=0.83\columnwidth,trim=45 30 0 0,clip=true]{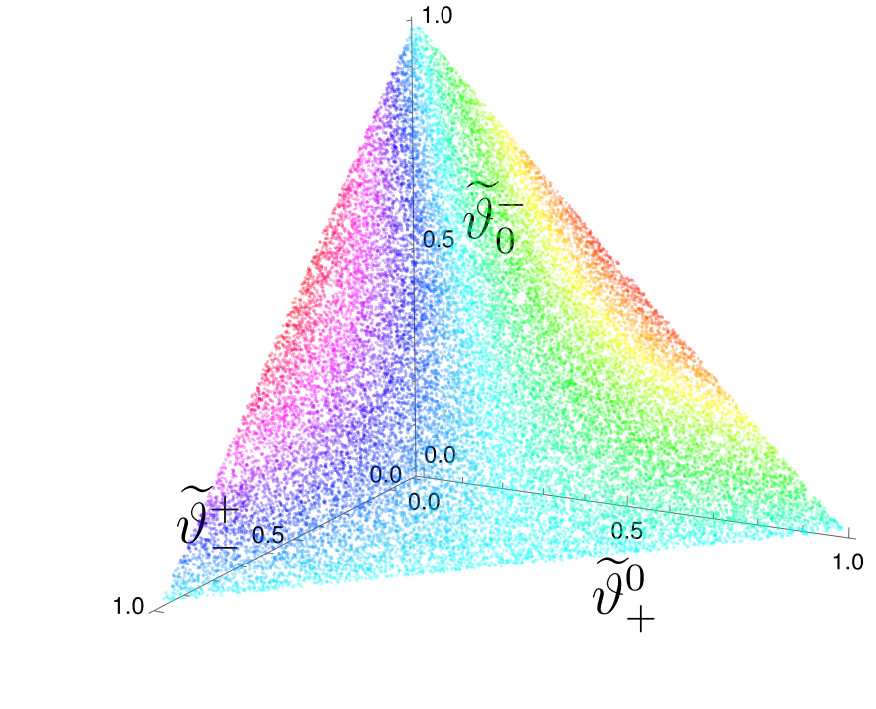}\includegraphics[width=0.17\columnwidth]{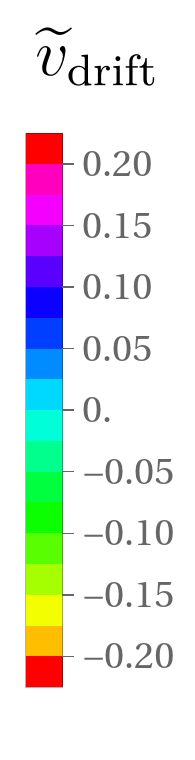}
\caption{The dimensionless $\widetilde{v}_\textrm{drift}=v_\textrm{drift}/v$ \eqref{vdrift-3cycle}, is represent as a color map on the space of the three dimensionless transitions rates $\{\widetilde{\vartheta}_{-}^{+},\widetilde{\vartheta}_{+}^{0},\widetilde{\vartheta}_{0}^{-}\}$, that form a 2-simplex (triangle). The diffusive case corresponds for the transitions rates that satisfy $\widetilde{v}_\textrm{drift}=0$.} 
\label{fig:vdrift-3cycles}
\end{figure}

\subsubsection{Diffusive transport}
Consider the cycle process where the transition rates $\{\vartheta_-^+,\vartheta_+^0,\vartheta_0^-\}$ determine the non-reciprocal dynamics. In this case the long-term diffusive behavior is observed when 
\begin{equation}\label{vdrift-3cycle}
v_\text{drift} = \frac{v}{\Theta^2} \vartheta_{0}^{-}\bigl(\vartheta_{-}^{+}-\vartheta_{+}^{0}\bigr)
\end{equation}
vanishes, i.e., whenever $\vt{-}{+}=\vt{+}{0}$. The dimensionless $\widetilde{v}_\textrm{drift}$ is shown in Fig.~\ref{fig:vdrift-3cycles}, for $2.5\times10^4$ random and uniformly sampled triads of dimensionless transitions rates $\{\widetilde{\vartheta}_-^+,\widetilde{\vartheta}_+^0,\widetilde{\vartheta}_0^-\}$, whose structure corresponds to a 2-simplex or triangle (see Sect.\ref{Sect:StationarySelfPropulsionSates}). The set of triads for which  $\widetilde{v}_\textrm{drift}=0$ is conspicuous. In this set, the long-term transport is diffusive with effective diffusion coefficient
\begin{figure}[h]
\includegraphics[width=\columnwidth,trim=5 10 0 0]{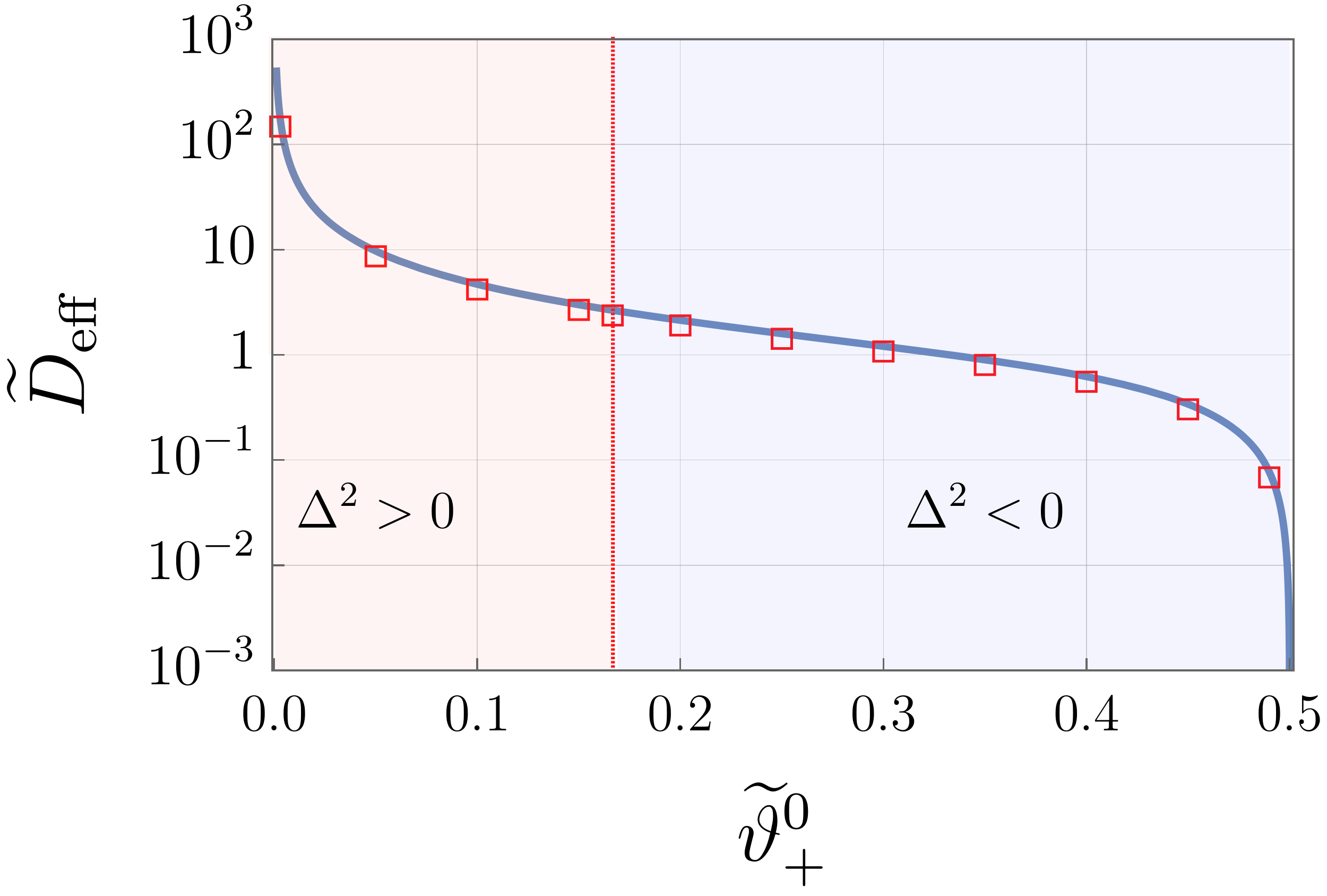}
\caption{The effective diffusion coefficient, $\widetilde{D}_\text{eff}$, as function of the transition rate $\widetilde{\vartheta}_{+}^{0}$ for diffusive transport ($v_\textrm{drift}=0$). The solid line indicates Eq.~\eqref{Eq:Dif_eff_3b} while the red squares mark the results obtained from numerical simulations. The vertical red line marks the crossover value $\widetilde{\vartheta}_{+}^{0}=1/6$, for which $\Delta^2=0$ and the diminishing rate of $D_\textrm{eff}$ is minimum.
}
\label{fig:Deff_3rates_vd0}
\end{figure}
\begin{subequations}
\begin{equation}\label{Eq:Dif_eff_3a}
   D_\textrm{eff}=v^2\dfrac{\vt{0}{-}}{\vt{+}{0}}\left(\dfrac{1}{\vt{+}{0}+2\vt{0}{-}}\right),
\end{equation}
where we have chosen $\vartheta_+^0$ and $\vartheta_0^-$ as independent parameters. The range of values of $D_\textrm{eff}$ is wide, giving control of the diffusive transport by properly tuning the transition rates. We can reduce expression \eqref{Eq:Dif_eff_3a} further by setting the inverse of  $\Theta=\vartheta_-^++\vartheta_+^0+\vartheta_0^-$ as time scale, and thus we can write as function of $\vartheta_+^0$ only, i.e.
\begin{equation}\label{Eq:Dif_eff_3b}
D_\textrm{eff}=v^2\frac{\Theta-2\vt{+}{0}}{\vt{+}{0}\bigl(2\Theta-3\vt{+}{0}\bigr)}, \quad 0<\vt{+}{0}<2\Theta.
\end{equation}    
\end{subequations}

The dependence of $\widetilde{D}_\textrm{eff}=D_\textrm{eff}/(v^2/\Theta)$ on $\widetilde{\vt{+}{0}}=\vartheta_+^0/\Theta$ is shown in Fig.~\ref{fig:Deff_3rates_vd0}, where it is observed a monotonous decrease from arbitrarily large values of $D_\textrm{eff}$ for  $\vt{+}{0}\approx0$, and diminishing to zero as $\vt{+}{0}\rightarrow\Theta/2$. In this case, the transition rate from the $v_+$ state to the resting state $v_0=0$ determines the diffusion coefficient. The large values of $D_\textrm{eff}$ are consequence of \emph{giant diffusion} effects \cite{LindnerNicola2008}, result of the rare transitions from self-propulsion state $v_+$ to $v_0=0$ and from $v_-$ to $v_+$. This make the initial occupation of the self-propulsion states $v_-$, $v_+$ to linger in those states, thus making the ballistic regime to persist times of the order of ${(\vt{+}{0}})^{-1}$ at which the crossover to diffusive dynamics appears. 

On the contrary, as $\vt{+}{0}=\vt{-}{+}\rightarrow\Theta/2$ we have $\vt{0}{-}\rightarrow0$, thus the transitions towards the state $v_0=0$ are far more frequent than those that takes the system out of this last one. This makes the initial occupation of the states $v_\pm$ to be depleted while the occupation of $v_0$ is increased, leading to a slowing down of the dynamics, and a small effective diffusion coefficient as consequence. 

The rate at which $D_\textrm{eff}$ diminishes as $\vt{+}{0}$ increases has a minimum at $\vt{+}{0}=\Theta/6$, this value corresponds to the one for which $\Delta^2$ \eqref{Delta2} vanishes, this values is marked in Fig.~\ref{fig:Deff_3rates_vd0}, which separates the region where $\Delta^2>0$ from the region where $\Delta^2<0$. It can be shown analytically that for the three-transition rates cycle process $-\Theta/4\le\Delta^2\le\Theta/12$. 

\begin{figure}
\includegraphics[width=\columnwidth, trim=0 10 0 0,clip=true]{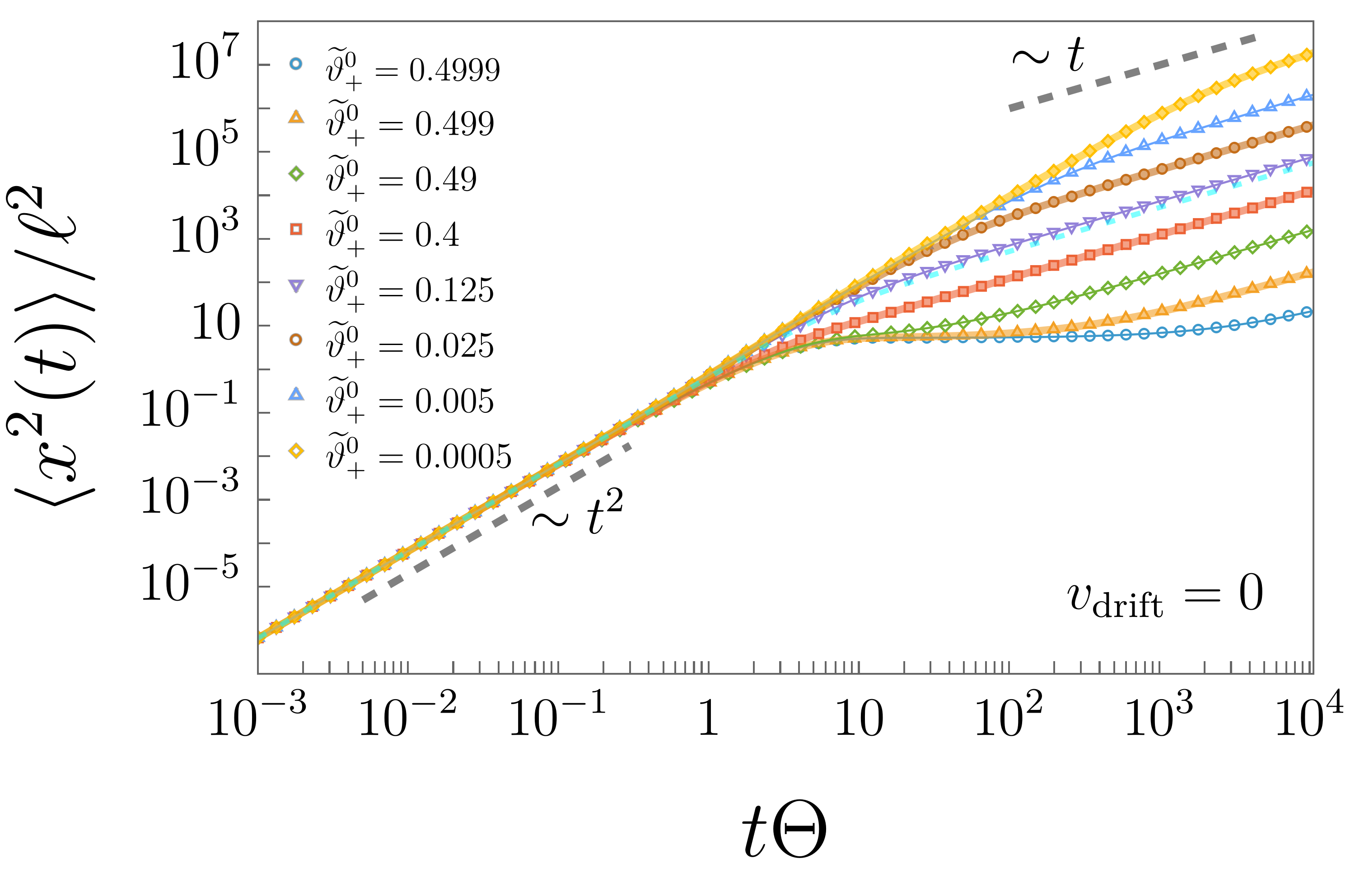}
\caption{The dimensionless MSD ($\ell=v/\Theta$) as function of the dimensionless time $t\Theta$ for different values of $\widetilde{\vartheta}_+^0$. The case $v_\textrm{drift}=0$ in the configuration of three transition rates $\{\vt{-}{+},\vt{+}{0},\vt{0}{-}\}$ is considered (see text). Symbols mark the results obtained from the ensemble average of $10^5$ trajectories obtained from numerical integration of Eq.~\eqref{Langevin}. Thin lines correspond to the evaluation of the corresponding analytical expressions.
}
\label{fig:MSD_Vd0_non-adj}
\end{figure}

These characteristics are corroborated by observing the time dependence of the MSD show in Fig.~\ref{fig:MSD_Vd0_non-adj}, where the dimensionless MSD is shown as function of the dimensionless time $\Theta t$ for $\widetilde{\vartheta}_+^0=0.0005,$ 0.005, 0.025, 0.125, 0.4, 0.49, 0.499 and 0.4999. The case of large $D_\textrm{eff}$ is clearly instanced by $\widetilde{\vartheta}_+^0=0.0005$ (yellow diamond symbols in Fig.~\ref{fig:MSD_Vd0_non-adj}) for which $\widetilde{D}_\textrm{eff}\approx9.9975\times10^2$; while the case of small diffusion is exemplified by $\widetilde{\vartheta}_+^0=0.4999$ (blue circle symbols in Fig.~\ref{fig:MSD_Vd0_non-adj}) with $\widetilde{D}_\textrm{eff}\approx7.997\times10^{-4}$. The MSD for $\vartheta_+^0=\Theta/6\approx0.6667$ is marked with cyan dots close to the curve with $\vt{+}{0}=0.125$ (purple down triangles). For all cases in the figure, symbols refer to the results obtained from numerical simulations while thin lines refer to the analytical expressions.  

\begin{figure}[h]
\includegraphics[width=0.95\columnwidth,trim=5 8 0 0,clip=true]{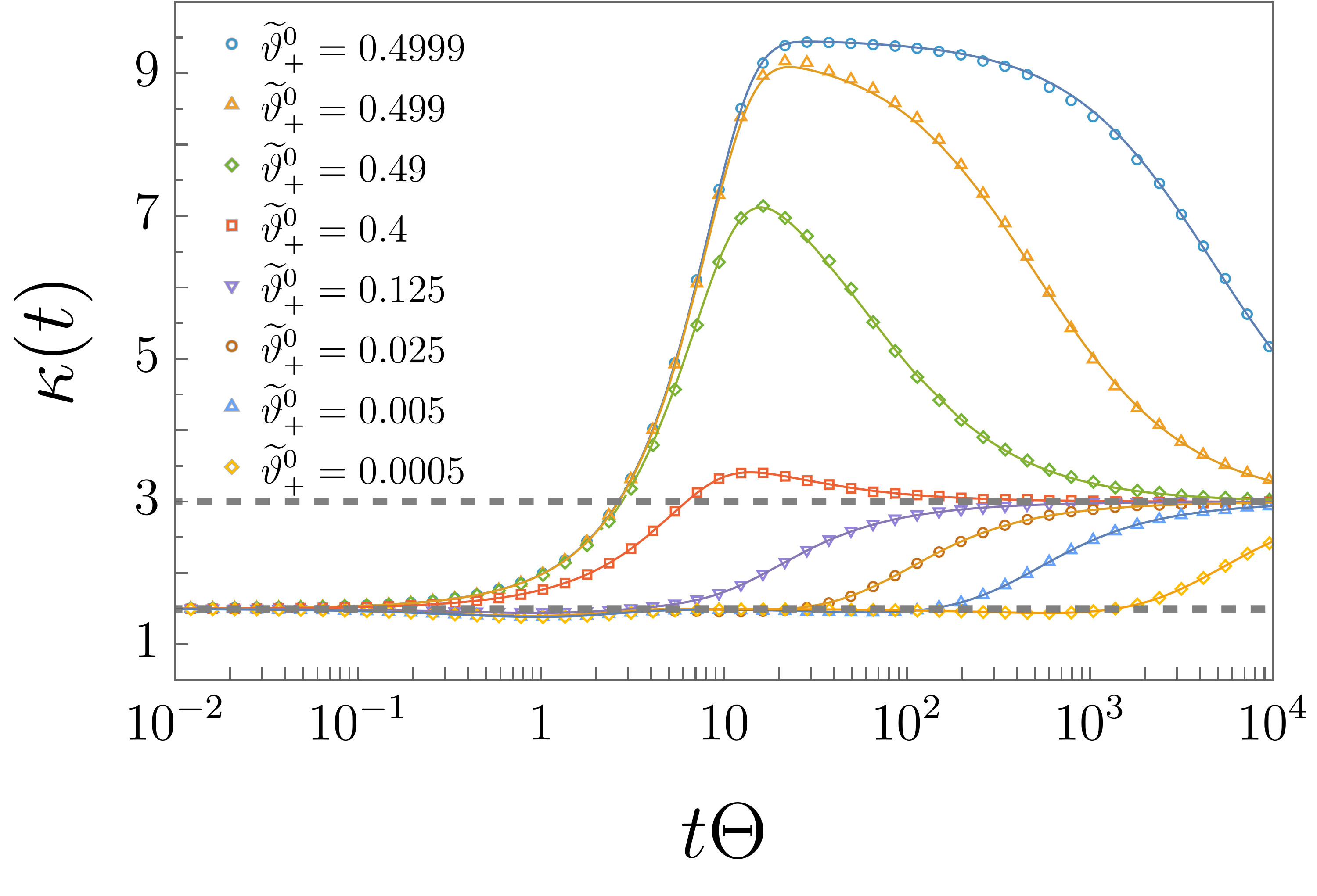}
\caption{The kurtosis of $p(x;t)$ \eqref{eq:kurtosis} as function of $t\Theta$ is shown for different values of $\widetilde{\vartheta}_+^0$. The diffusive case is considered ($v_\textrm{drift}=0$) in the cyclic configurations of three transition rates \eqref{eq:Cycle-3rates}. In the short time limit, where transitions between self-propulsion states plays no relevant role, the value 3/2 characterizes the initial distribution (three pulses), the time evolution is diverse determined by $\vt{+}{0}$. In the long-time regime the kurtosis reaches asymptotically the characteristic values for a Gaussian 3 (see Eq.~\eqref{eq:Gaussian}).  
}
\label{fig:kurtosis_non-adj_rates}
\end{figure}

Finally, in Fig.~\ref{fig:kurtosis_non-adj_rates} the time dependence of the kurtosis of $p(x;t)$ is shown for the chosen values of $\vt{+}{0}$ (symbols mark the results obtained from our numerical simulations while thin-solid lines refer to their corresponding analytical expressions, not shown in the paper). For the values of $\vt{+}{0}<\Theta/6$, which make $\Delta^2>0$, the kurtosis is bounded above by the value of a Gaussian distribution 3, values which is approached asymptotically. This characteristic implies a that $p(x;t)$ evolves ``compactly'' dispersion plays a small role. In contrast $\vt{+}{0}>\Theta/6$, which implies $\Delta^2<0$ (typical case), the kurtosis overpasses the value 3 relaxing to the asymptotically to it. In this case, the slowing down induced by the increased of the probability of being at the state $v_0=0$, makes the distribution of the positions to have ``outliers'' which makes the kurtosis to have values larger than 3 \cite{WestfallAmStat2014}. 

\subsection{Cycle of four transition rates}
\begin{figure}
\centering
\includegraphics[width=0.87\columnwidth]{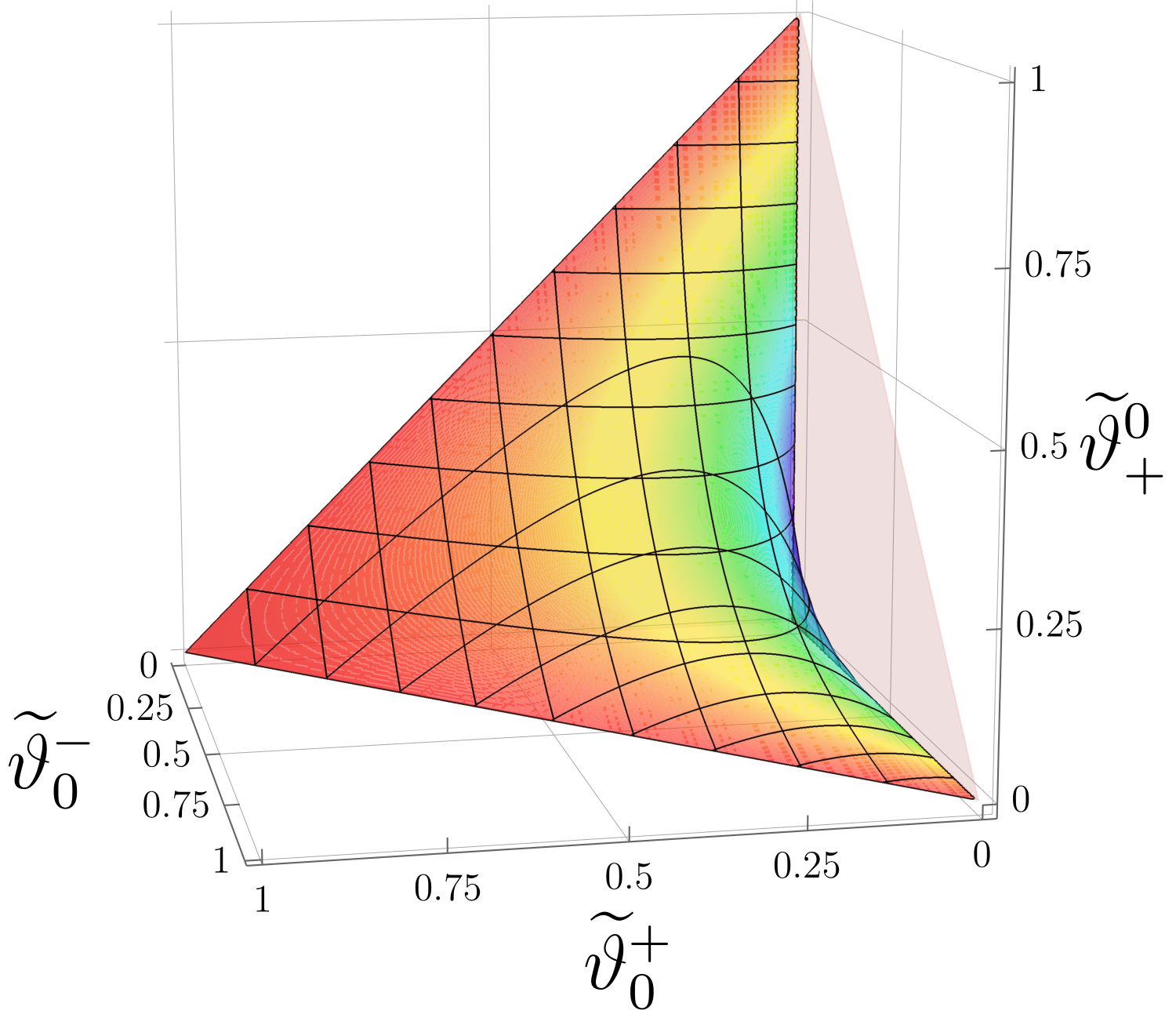}\includegraphics[width=0.13\columnwidth]{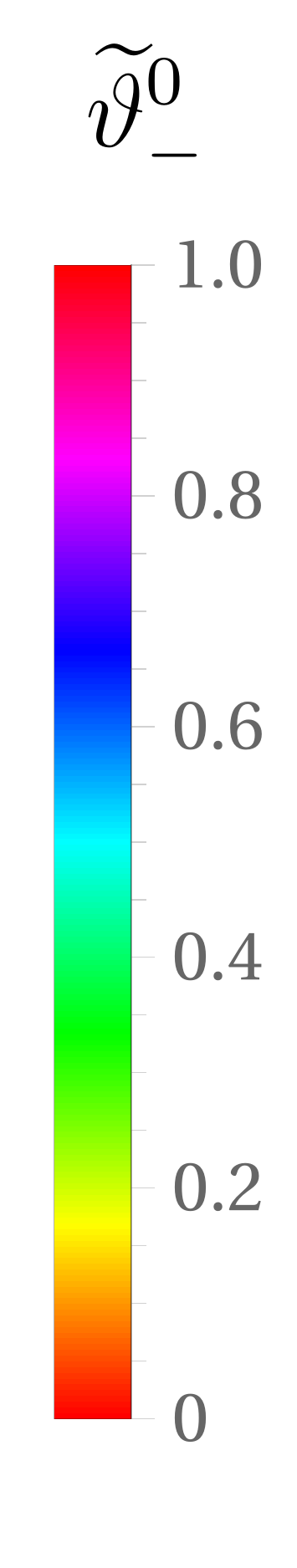}
\caption{The manifold defined by vanishing of the drift velocity $\widetilde{v}_\text{drift}=\bigl(\widetilde{\vartheta}_-^0\widetilde{\vartheta}_0^+-\widetilde{\vartheta}_+^0\widetilde{\vartheta}_0^-\bigr)=0$ is shown in the parameter space  $\bigl\{\widetilde{\vartheta}_-^0,\widetilde{\vartheta}_0^+,\widetilde{\vartheta}_+^0,\widetilde{\vartheta}_0^-\bigr\}$. The color of each point of the surface gives the value of $\widetilde{\vartheta}_-^0$ according to the bar.} 
\label{fig:Manifold-4rates}
\end{figure}

The second particular case of our study consists of
four transition rates such that the long transitions $\vt{+}{-}$, $\vt{-}{+}$ are neglected (see bottom panel of Fig.~\ref{fig:Transition_non-adj_rates}),
\begin{equation}
\label{4-cycle}
v_-\overset{\vartheta_-^0}{\underset{\vartheta_0^-}{\longleftrightarrow}} v_0
\overset{\vartheta_0^+}{\underset{\vartheta_+^0}{\longleftrightarrow}} v_+.
\end{equation}
We are interested in the configurations of the four transition rates chosen, that leads to diffusive behavior, i.e., we focus on the manifold in the space defined by $\{\vt{-}{0},\vt{0}{+},\vt{+}{0},\vt{0}{-}\}$ such that
\begin{subequations}
\begin{equation}\label{eq: vdrift_4rates}
    v_{\text{drift}} = \dfrac{v}{\Theta^2}\bigl(\vartheta_{0}^{+} \vartheta_{-}^{0} - \vartheta_{0}^{-} \vartheta_{+}^{0}\bigr)=0,
\end{equation}
and thus 
\begin{equation}
\label{eq:4rates_vdrift0}
\vt{0}{+}\vt{-}{0}=\vt{0}{-}\vt{+}{0}. 
\end{equation}
\end{subequations}
This manifold is shown in Fig.~\ref{fig:Manifold-4rates}, where each colored point on the surface gives the values of the dimensionless transition rate $\widetilde{\vartheta}_-^0$ in the space of the other three dimensionless transitions rates $\widetilde{\vartheta}_0^+$, $\widetilde{\vartheta}_+^0$ and 
$\widetilde{\vartheta}_0^-$, where the transitions rates are made dimensionless with $\Theta=\vt{-}{0}+\vt{0}{+}+\vt{+}{0}+\vt{0}{-}$.

\subsubsection{Diffusive transport}
\begin{figure}
\centering
\includegraphics[width=\columnwidth]{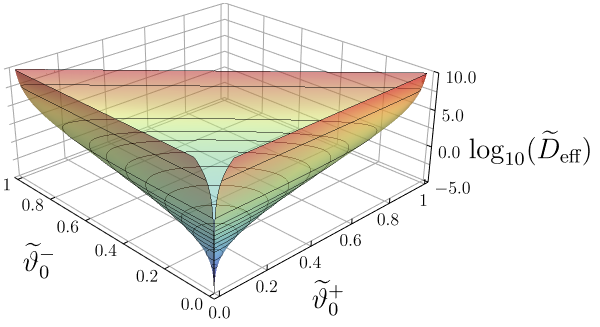}
\caption{The dimensionless effective diffusion coefficient $\widetilde{D}_{\text{eff}}$ is shown as function of $\vt{0}{-},\vt{0}{+}$ as given by Eq.~\eqref{eq:Deff_4rates1-2}.
}
\label{fig:Deff_4rates}
\end{figure}

In this case, the effective diffusion is given by 
\begin{equation}
D_\textrm{eff}=v^2\frac{\vt{0}{-}+\vt{0}{+}}{\vt{0}{-}\vt{+}{0}+\vt{-}{0}\vt{+}{0}+\vt{0}{+}\vt{-}{0}},
\end{equation}
and due to the constriction \eqref{eq:4rates_vdrift0} and to the definition of $\Theta$, $D_\textrm{eff}$ can be expressed as a function of six different pairs among the transitions rates $\{\vt{-}{0},\vt{0}{+},\vt{+}{0},\vt{0}{-}\}$. We chose the pair of transitions rates that start at the state $v=0$, i.e. $\vt{0}{-},\vt{0}{+}$. Thus
\begin{equation}\label{eq:Deff_4rates1-2}
D_\textrm{eff}(\vartheta_{0}^{-}, \vartheta_{0}^{+}) = v^2\,\dfrac{(\vartheta_{0}^{-} + \vartheta_{0}^{+})^3}{\vartheta_{0}^{-} \vartheta_{0}^{+} \left[\Theta^2 - (\vartheta_{0}^{-} + \vartheta_{0}^{+})^2\right]},
\end{equation}
from which several limiting behaviors can be identified as can be observed in Fig.~\ref{fig:Deff_4rates}. First notice the appearance of giant diffusion when either \(\vartheta_{0}^{-} \to 0\) or \(\vartheta_{0}^{+} \to 0\), and when \(\vartheta_{0}^{-} + \vartheta_{0}^{+} \to \Theta\), for which $D_\textrm{eff}$ diverges. These results suggest that, in the first two cases, particles tend to dwell in one active state leading to a prolonged directed motion that enhances the effective diffusion. On the contrary, when both $\vartheta_0^-$, $\vartheta_0^+\rightarrow0$, the rest state dominates the transient period leading to small values of $D_\textrm{eff}$.  

\begin{figure}
\centering
\includegraphics[width=\columnwidth,trim=0 20 0 0]{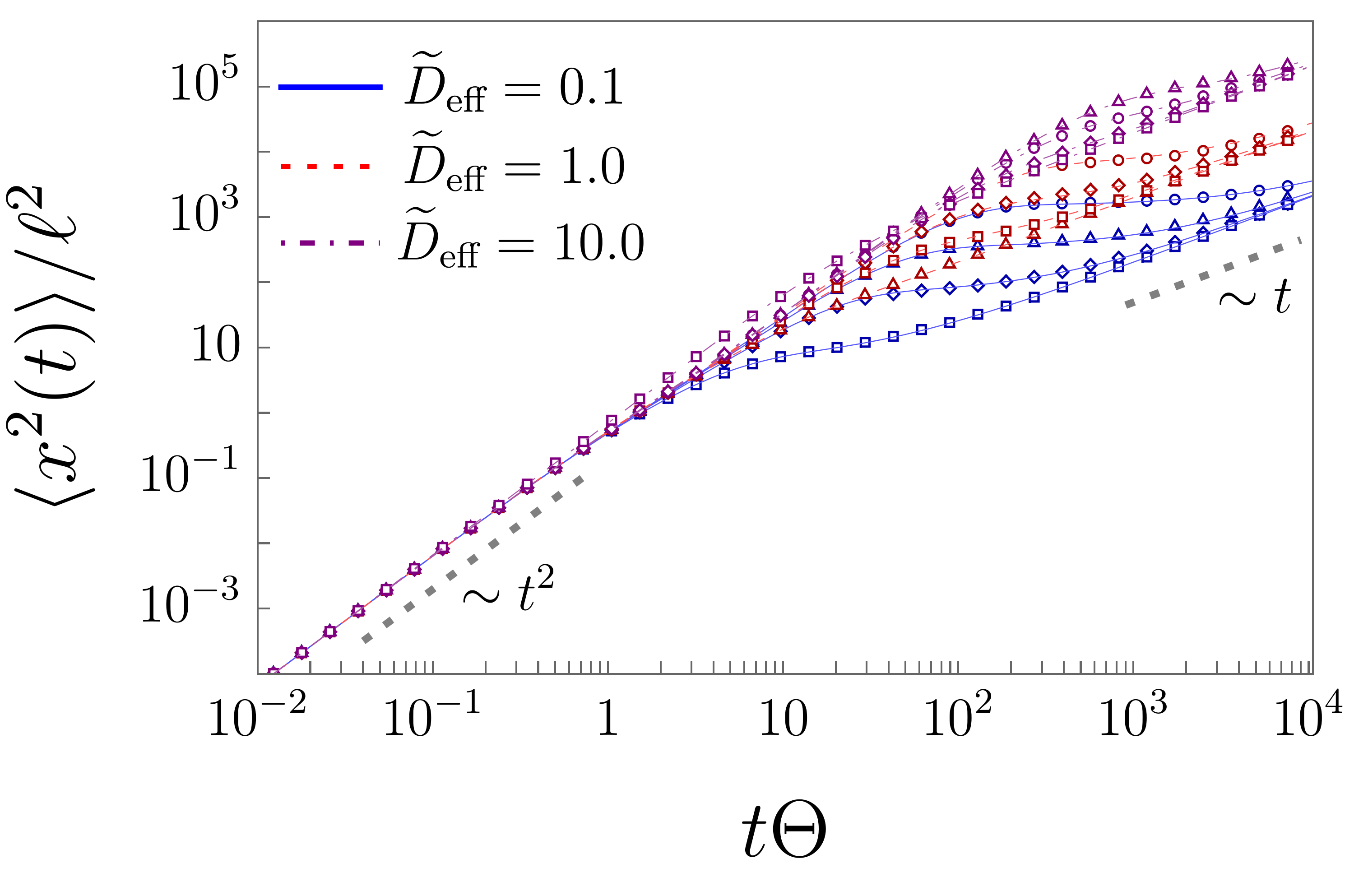}
\caption{The time dependence of the mean square displacement (diffusive regime $v_\textrm{drift}=0$) for 4 distinct configuration $\{ \widetilde{\vartheta}_-^0,\widetilde{\vartheta}_0^+,\widetilde{\vartheta}_+^0,\widetilde{\vartheta}_0^-\}$ is shown. Distinct configurations are chosen according to Table~\ref{table:rates_4states}  along the level curves $D_\textrm{eff}=0.1$ (solid-blue lines), 1.0 (dashed-red lines) and 10.0 (dash-dotted-purple lines) of Fig.~\ref{fig:Deff_4rates};  $\widetilde{\vartheta}_-^0$ (squares), $\widetilde{\vartheta}_0^+$ (diamonds), $\widetilde{\vartheta}_+^0$ (up triangle), and $\widetilde{\vartheta}_0^-$ (circles).}
\label{fig:sample_msd_4r}
\end{figure}

\begin{table}[h!]
\centering
\caption{Set of selected dimensionless transition rates $\{\widetilde{\vartheta}_-^0,\widetilde{\vartheta}_0^+,\widetilde{\vartheta}_+^0,\widetilde{\vartheta}_0^-\}$ used for the computation of the MSD curves in Fig.~\ref{fig:sample_msd_4r}. Each block corresponds to a fixed value of the effective diffusion coefficient $\widetilde{D}_\textrm{eff}$.}
\begin{tabular}{ccccc}
$\widetilde{D}_\textrm{eff}$ &
$\widetilde{\vartheta}_-^0$ &
$\widetilde{\vartheta}_0^+$ &
$\widetilde{\vartheta}_+^0$ &
$\widetilde{\vartheta}_0^-$ \\
\specialrule{0.12em}{0em}{0em}   
\multirow{4}{*}{0.1}
    & 0.0208 & 0.002  & 0.9771 & 0.00004 \\
    & 0.0435 & 0.004  & 0.9522 & 0.0001 \\
    & 0.0975 & 0.008  & 0.8935 & 0.0008 \\
    & 0.6339 & 0.008  & 0.3432 & 0.0147 \\
\specialrule{0.06em}{0em}{0em}
\multirow{4}{*}{1.0}
    & 0.1010 & 0.01   & 0.9797 & 0.0001 \\
    & 0.7359 & 0.02   & 0.1351 & 0.1089 \\
    & 0.0204 & 0.02   & 0.9591 & 0.0004 \\
    & 0.0419 & 0.04   & 0.9161 & 0.0018 \\
\specialrule{0.08em}{0em}{0em}
\multirow{4}{*}{10.0}
    & 0.0048 & 0.05   & 0.9449 & 0.0002 \\
    & 0.0029 & 0.03   & 0.9669 & 0.00009 \\
    & 0.0092 & 0.1    & 0.8896 & 0.0010 \\
    & 0.0818 & 0.45   & 0.1    & 0.3681 \\
\specialrule{0.12em}{0em}{0em} 
\end{tabular}
\label{table:rates_4states}
\end{table}

We want to point out that 
even in the same level curve defined by a specific value of $D_\textrm{eff}$ on the surface of Fig.~\ref{fig:Deff_4rates}, the time dependence of the mean squared displacement can vary significantly with the transition rates. In Fig.~\ref{fig:sample_msd_4r} we show the time dependence of the MSD for
three representative values of $D_{\text{eff}}$, namely 0.1 (solid-blue line), 1.0 (dashed-red), and 10.0 (dash-dotted purple line). For each of these values, four MSD curves are shown (differentiated by squares, diamonds, up-triangles and circles), each corresponding to a different point uniformly sampled along the respective level curve of the $D_{\text{eff}}$ surface (see the values chosen at Table~\ref{table:rates_4states}). It is clear, despite sharing the same effective diffusive coefficient, the MSD curves differ in their short- and intermediate-time regimes. In particular, configurations of transfer rates that give higher $D_{\text{eff}}$, sustain the ballistic regime for longer durations, and the crossover to the diffusive regime occurs at different characteristic times depending on the point chosen along the contour.

\section{Conclusion and final remarks}
\label{Sect:Conclusions}
In this work we have investigated the transport properties of active particles undergoing run-and-tumble dynamics among three self-propulsion velocity states $\{-v, 0, v\}$, incorporating explicitly non-reciprocal transition rates between them. Our analytical and numerical study reveals that the breaking of microscopic reversibility on the self-propulsion dynamics fundamentally enriches the emergent transport regimes. Despite the apparent complexity of the system, we derived exact expressions, in the Fourier-Laplace domain, for the marginal probability density of the particle positions, from which characteristic transport quantities as the long-term effective drift velocity and the effective diffusion coefficients were identified.

Importantly, we have elucidated  the manifold, in the transition-rate space, where long-time diffusive behavior is observed ($v_\textrm{drift}=0$), despite the absence of microscopic reversibility. This results allows for the design of the system configurations to control the transport properties and the macroscopic long-term properties that emerge from fundamentally irreversible dynamics at the microscopic level. Remarkably, we uncovered the emergence of giant diffusion, where the effective diffusion coefficient diverges when the transition rates are tuned to prolong the ballistic regime. 

The geometry of the parameter space, represented as a 5-simplex of the dimensionless transition rates, not only organizes the possible dynamics, but also allows for a global exploration of the parameter space by analyzing the distribution of the values of  characteristic quantities when generating a sample of uniformly  six-tuples of transitions rates chosen at random. This method reveals universal features such as entropy extrema and symmetry-induced diffusion. 

Altogether, our results emphasize the pivotal role of transition-rate asymmetry in shaping transport in active systems and provide a theoretical framework for engineering tunable propagation by controlling internal dynamics. Our results demonstrate a clear transition between ballistic, diffusive, and anomalous regimes, consistent with previously reported features such as super-diffusion and non-Gaussian statistics. Furthermore, the ability to tune the effective diffusion through the irreversibility of internal transition rates broadens the range of emergent transport behaviors captured by the model. The model and methodology presented here may be extended to more intricate scenarios involving confinement, interactions, or heterogeneous media, opening new avenues for the study of non-equilibrium transport.

\begin{acknowledgments}
This work was supported by UNAM-PAPIIT IN112623. 
J.C.R.R.C. acknowledges doctoral scholarship 778638 from SECIHTI (previously CONAHCYT).
\end{acknowledgments}

\appendix

\section{Generalized current for run-and-tumble active motion}
\label{app:InitialConditions}
We consider the generalisation of Eqs. \eqref{3stat-masteq} to the case of arbitrary velocities $\{v_-,v_0,v_-\}$. The master equations in Fourier domain are given by:
\begin{subequations}\label{3StatesME}
\begin{equation}
\label{eq:MasterEqs-Gral}
\partial_t\hat{p}_-+ \textrm{i}k v_{-}\, \hat{p}_{-}=\vartheta_+^-\,\hat{p}_{+}+\vartheta_0^-\,\hat{p}_0-(\vartheta_{-}^{0}+\vartheta_-^+)\,\hat{p}_-,    
\end{equation}
\begin{equation}
\partial_t\hat{p}_0+ \textrm{i}k v_{0}\, \hat{p}_{0}=\vartheta^{0}_{-}\,\hat{p}_-+\vartheta^{0}_{+}\,\hat{p}_{+}-(\vartheta_{0}^{-}+\vartheta_{0}^{+})\, \hat{p}_0,
\end{equation}
\begin{equation}
\partial_t\hat{p}_++ \textrm{i} k v_{+}\, \hat{p}_+=\vartheta_{-}^{+}\,\hat{p}_-+\vartheta_{0}^{+}\,\hat{p}_0-(\vartheta_{+}^{-}+\vartheta_{+}^{0})\, \hat{p}_{+}.
\end{equation}
\end{subequations}

The marginal probability density, $p(x;t)$, of finding a particle at the position $x$ at time $t$ independent of the velocity states and its corresponding probability current $j(x;t)$ are, respectively, 
\begin{subequations}
\begin{align}
 p(x;t)&\equiv p_{-}(x;t) + p_0(x;t) + p_+(x;t),\\
j(x;t)&\equiv v_- p_{-}(x;t) + v_0 p_0(x;t) + v_+ p_+(x;t). \label{eq:gral_current}
\end{align}
\end{subequations}
These satisfy the continuity equation $\partial_t p(x;t)+\partial_x j(x,t)=0$. In Fourier domain we have 
\begin{subequations}
\begin{align}
 \hat{p}(k;t)&\equiv \hat{p}_{-}(k;t) + \hat{p}_0(k;t) + \hat{p}_+(k;t)\\
\hat{j}(x;t)&\equiv v_- \hat{p}_{-}(k;t) + v_0 \hat{p}_0(k;t) + v_+ \hat{p}_+(k;t)\label{gral-JFourier}
\end{align}
\end{subequations}
that satisfy $\partial_t \hat{p}(k;t)+\textrm{i}k\hat{j}(k,t)=0$. Thus
\begin{align}
\partial_t\hat{p}^{(0)}(k)=&\partial_t\hat{p}(k;t)\big\vert_{t=0}\nonumber\\
=&-\textrm{i}k\hat{j}^{(0)}(k)\nonumber\\
=&-\textrm{i}k\bigl[ v_- \hat{p}_-^{(0)}(k) + v_0 \hat{p}_0^{(0)}(k) + v_+ \hat{p}_+^{(0)}(k)\bigr]
\end{align}
where we have used \eqref{gral-JFourier} and the definition $\hat{j}^{(0)}(k)=\hat{j}(k;t)\vert_{t=0}$. In the case considered in this work $v_+=v=-v_-$ and $v_0=0$
\begin{equation}\label{IniCurrent}
\partial_t\hat{p}^{(0)}(k)=-\textrm{i}k\bigl[v\bigl(\hat{p}_+^{(0)}(k)-\hat{p}_-^{(0)}(k)\bigr)\bigr].
\end{equation}
This last expression vanishes for the initial conditions chosen \eqref{eq:IniDistributions}.

Since $\hat{j}^{(0)}(k)=\int_{-\infty}^{\infty} dx\, e^{-\textrm{i}kx}j(x;t)\big\vert_{t=0}$ we have that
\begin{equation}\label{TotalCurrent}
\hat{j}^{(0)}(k)\Big\vert_{k=0}=\int_{-\infty}^\infty dx\, j(x;t)\big\vert_{t=0}=J^{(0)}
\end{equation}
corresponds to the total initial current. Related to the calculation of the MSD in Sect.~\ref{sect:MSD}, we consider $\bigl[\textrm{i}\partial_k\bigl(\partial_t\hat{p}^{(0)}(k)\bigr)\bigr]_{k=0}$, which by use of the continuity equation we get $\bigl[\partial_k \bigr(k\, \hat{j}^{(0)}(k)\bigr)\bigr]_{k=0}$, after rearranging terms we have that
\begin{equation}
\Bigl[\textrm{i}\partial_k\bigl(\partial_t\hat{p}^{(0)}(k)\bigr)\Bigr]_{k=0}=J^{(0)}.
\end{equation}

After taking the partial derivative with respect to $k$ again to the last expression, we have 
\begin{align}
\partial_{kk}\left[\partial_t \hat{p}^{(0)}(k)\right]&=-2\textrm{i} \left[\partial_k \hat{j}^{(0)}(k) \right]_{k=0} \nonumber \\
&=-2v \left(\textrm{i}\partial_k \hat{p}_+^{(0)}(k) - \textrm{i}\partial_k \hat{p}_-^{(0)}(k) \right) \nonumber \\
&=-2v\left(\bigl\langle x(0)\bigr\rangle_+ - \bigl\langle x(0)\bigr\rangle_- \right)\nonumber\\
&:=-\gamma,\label{eq:gamma}
\end{align}
where the parameter $\gamma$ quantifies the initial spatial asymmetry among the velocity states occupation and vanishes for the initial distributions \eqref{eq:IniDistributions}.

On the other hand
\begin{align}
\partial_{tt}\hat{p}^{(0)}(k)&=\partial_{tt}\hat{p}(k;t)\big\vert_{t=0}\nonumber\\
&=-\textrm{i}k\partial_t\hat{j}(k;t)\big\vert_{t=0},\label{DotIniCurrent}
\end{align}
where the partial derivative with respect to time of $\hat{j}(k;t)$ is given by 
\begin{equation}\label{eq:D1_j}
\partial_t\hat{j}(k;t)= v_-\partial_t\hat{p}_-(k;t) + v_0 \partial_t\hat{p}_0(k;t) + v_+\partial_t\hat{p}_+ (k;t).
\end{equation}
After using the master equations with arbitrary velocities \eqref{eq:MasterEqs-Gral} we have that
\begin{multline}
-\textrm{i} k \partial_t \hat{j}(k;t) = \\ 
\left[-k^2 v^2_{-} - \textrm{i}k (\vt{-}{0}v_0 + \vt{-}{+} v_+ - (\vt{-}{0}+\vt{-}{+}) v_{-} \right]\hat{p}_-(k;t)+\\
\left[-k^2 v^2_{0} - \textrm{i}k (\vt{0}{-}v_- + \vt{0}{+} v_+ -(\vt{0}{-} + \vt{0}{+}) v_0) \right]\hat{p}_0(k;t)+\\
\left[-k^2 v^2_{+} - \textrm{i}k (\vt{+}{-} v_- + \vt{+}{0} v_0 - (\vt{+}{-}+\vt{+}{0}) v_+) \right]\hat{p}_+(k;t),
\end{multline}
which after taking the particular values of the velocities $v_-=-v,\, v_0=0$, and $v_+=v$, and after evaluating at $t=0$, we have 
\begin{multline}\label{2partialTimeP0}
 \partial_{tt}\hat{p}^{(0)}(k)=-\textrm{i} k \partial_t \hat{j}(k;t)\bigl|_{t=0} \\ 
=\left[-k^2 v^2 - \textrm{i}k v (\vt{-}{0} + 2\vt{-}{+}) \right]\hat{p}_-^{(0)}(k) \\
+\textrm{i}k v\,  (\vt{0}{-}-\vt{0}{+})\hat{p}_0^{(0)}(k)\\
+\left[-k^2 v^2 + \textrm{i}k v (\vt{+}{0} + 2 \vt{+}{-}) \right]\hat{p}_+^{(0)}(k).
\end{multline}

After taking the partial derivative with respect to $k$, and evaluating at $k=0$ we get
\begin{align}
\partial_k[\partial_{tt}\hat{p}^{(0)}(k)]\Bigr\vert_{k=0}&=-\textrm{i}v\bigl[(\vartheta_-^0+2\vartheta_-^+)\hat{p}_-^{(0)}(0)\nonumber\\
&+(\vartheta_0^+-\vartheta_0^-)\hat{p}_0^{(0)}(0)\nonumber\\&
-(\vartheta_+^0+2\vartheta_+^-)\hat{p}_+^{(0)}(0)\bigr]\nonumber\\
&\coloneqq-\textrm{i}\,\alpha,\label{eq:alpha}
\end{align}
where we have introduced $\alpha$, which has units of length per time squared. For the initial distributions \eqref{eq:IniDistributions} considered in this paper, these are $k$-independent $\hat{p}^{(0)}_-(k)\; \hat{p}^{(0)}_0(k)\; \hat{p}^{(0)}_+(k)=\frac{1}{3}$ and $\alpha=\frac{v}{3}[(\vartheta_-^0-\vartheta_0^-)+(\vartheta_0^+-\vartheta_+^0)+2(\vartheta_-^+-\vartheta_+^-)]$ for this case.

Finally, computing the second partial derivative in Eq.~\eqref{2partialTimeP0}, we obtain:
\begin{align}
\partial_{kk}[\partial_{tt}\hat{p}^{(0)}(k)]\Bigr\vert_{k=0} & = -2v^2 \left(\hat{p}_{-}^{(0)}(0)+\hat{p}_+^{(0)}(0) \right) \nonumber\\
&\quad - 2v \bigl[(\vt{-}{0}+2\vt{-}{+}) \bigl\langle x(0)\bigr\rangle_- \nonumber\\
&\quad + (\vt{0}{-}-\vt{0}{+})\bigl\langle x(0)\bigr\rangle_0 \nonumber \\ 
&\quad - (\vt{+}{0}+2\vt{+}{-})\bigl\langle x(0)\bigr\rangle_+ \bigr]\nonumber\\
&\coloneqq-\beta\label{def:beta}, 
\end{align}
which has units of velocity squared. Notice that in contrast to  $\gamma$ \eqref{eq:gamma} and $\alpha$ \eqref{eq:alpha}, for the initial distributions \eqref{eq:IniDistributions} $\beta$ is finite, even in the case of detailed-reciprocal transitions ($\vartheta_n^m=\vartheta$ for all $n,m\in\mathbb{S}$). $\beta$ vanishes if initial pulses at the origin such that $p_-^{(0)}(k)$, $p_-^{(0)}(k)=0$, $p_0^{(0)}=1$.

\section*{Data Availability}

The raw and processed data that support the findings of this study are openly available in Zenodo at [10.5281/zenodo.15875313], reference number.

\bibliography{References}
\bibliographystyle{apsrev}

\end{document}